\documentclass[iop]{emulateapj}
\usepackage{apjfonts}
\usepackage{times} 
\usepackage{graphicx}
\usepackage{rotating}
\usepackage{array}
\usepackage{epsfig}
\usepackage{amsmath}
\usepackage{natbib}
\def\mbh{\ifmmode{{\mathrm M}_{bh}\,}\else{M$_{bh}$\,}\fi}
\def\msigma{\ifmmode{{\mathrm M}_{bh}-{\sigma}\,}\else{M$_{bh}- \sigma$\,}\fi}
\def\msun{\ifmmode{{\mathrm M}_{\odot}}\else{M$_{\odot}$}\fi} 
\def\kms{\ifmmode{{\mathrm{km \, s^{-1}}}}\else{${\mathrm{km \, s^{-1}}}$}\fi}

\def\zmax{Z_{\rm max}}
\def\rperi{r_{\rm peri}}
\def\rapo{r_{\rm apo}}
\def\tform{t_{\rm form}}

\defcitealias{valluri_etal_10}{V10}
\defcitealias{valluri_etal_12}{V12}
\defcitealias{deb_etal_08}{D08}
\defcitealias{BT}{BT08}

\shorttitle{Halo orbits in a cosmological disk}
\shortauthors{Valluri et al. }

\begin{document}
\title{Halo orbits in cosmological disk galaxies: tracers of formation history}

\bigskip

\author{
Monica Valluri\altaffilmark{1},
Victor P. Debattista\altaffilmark{2},
Gregory S. Stinson\altaffilmark{3},
Jeremy Bailin\altaffilmark{4},
Thomas R. Quinn\altaffilmark{5},
H. M. P. Couchman\altaffilmark{6},
James Wadsley \altaffilmark{6}
}
\altaffiltext{1}{Department of Astronomy, University of Michigan, Ann Arbor, MI 48109, {mvalluri@umich.edu}}
\altaffiltext{2}{Jeremiah Horrocks Institute, University of Central Lancashire,  Preston, PR1 2HE, UK}
\altaffiltext{3}{Max-Planck-Institut f\"ur Astronomie, K\"onigstuhl 17, 69117, Heidelberg, Germany}
\altaffiltext{4}{Department of Physics and Astronomy, University of Alabama, Box 870324, Tuscaloosa, AL 35487-0324}
\altaffiltext{5}{Astronomy Department, University of Washington, Box 351580, Seattle, WA 98195-1580, USA}
\altaffiltext{6}{Department of Physics and Astronomy, McMaster University, Hamilton Ontario, L8S 4M1, Canada}

\journalinfo{Astrophysical Journal}

%\accepted{\today} 
\submitted{Submitted:Nov 14, 2012, Accepted: Feb 28, 2013}

\begin{abstract} 
We analyze the orbits of stars and dark matter particles in the halo
of a disk galaxy formed in a  cosmological hydrodynamical
simulation. The halo is oblate within the inner $\sim 20$~kpc and
 triaxial beyond this radius. About 43\% of orbits are
short axis tubes - the rest belong to orbit families that
characterize triaxial potentials (boxes, long-axis tubes and chaotic
orbits), but their shapes are close to axisymmetric. We find no
evidence that the self-consistent distribution function of the nearly
oblate inner halo is comprised primarily of axisymmetric short-axis tube
orbits.  Orbits of all families, and both types of particles are highly eccentric with mean
eccentricity $\gtrsim 0.6$.  We find that randomly selected samples of
halo stars show no substructure in ``integrals of motion''  space.
However individual accretion events can be clearly identified in plots of
metallicity versus formation time.  Dynamically young tidal debris is
found primarily on a single type of orbit. However, stars associated with older
satellites become chaotically mixed during the formation process
(possibly due to scattering by the central bulge and disk, and
baryonic processes), and appear on all four types of orbits.  We find that the tidal debris in cosmological hydrodynamical simulations experiences significantly more chaotic
evolution than in collisionless simulations, making it much harder to identify individual progenitors using phase space coordinates alone. However by combining information on stellar ages and chemical abundances with the orbital properties of halo stars in the underlying self-consistent potential, the identification of progenitors is likely to be possible.

\end{abstract}


\vspace{1.cm}

\section{Introduction}
\label{sec:intro}

In the picture proposed by \citet{SZ78} the stellar halo of the Milky
Way (hereafter MW) formed from the merger of protogalactic fragments
consisting of dark matter (hereafter DM), gas, and the first
generations of stars which formed in high density peaks in the early
Universe. An extended period of ``late infall'' and tidal shredding of
satellite galaxies has continued to build up the halo even at the
present time.  In the past decade, resolved-star surveys have provided
strong observational evidence in support of this picture.  Numerous
coherent stellar tidal streams \citep{newberg_etal_02, yanny_etal_03,
  belokurov_etal_06field} and tens of ultra-faint dwarf spheroidal
galaxies \citep{willman_etal_05ursa, zucker_etal_06canes,
  belokurov_etal_07} -- possibly progenitors of the tidal streams --
have been discovered in large surveys. Their discovery is broadly
consistent with the predictions of galaxy formation simulations in the
cold dark matter (hereafter $\Lambda$CDM) paradigm
\citep{bullock_johnston_05, bell_etal_08}.

In this paper we focus on the orbital properties of halo stars and
dark matter particles in a disk galaxy produced in an $N$-body+SPH
simulation run in a fully cosmological context \citep[part of the
  McMaster Unbiased Galaxy Simulations, hereafter
  MUGS]{stinson_etal_10}. The MUGS sample consists of 16 galaxies
(both ellipticals and spirals) with host halo masses of $\sim
10^{12}$~\msun\ selected in an unbiased way and resimulated with
volume renormalization techniques. The MUGS simulation has higher mass
resolution and smaller spatial gravitational softening than recent
simulations which have examined similar issues, but it still suffers
from the over cooling problem that has plagued most cosmological
simulations \citep[however this situation has recently improved,
  e.g.][]{governato_etal_10, guedes_etal_11, zemp_etal_12,
  stinson_etal_12}. The subject of our investigation is the most
disk-dominated massive spiral galaxy in the MUGS sample, which happens
to have a mass, and radial scale length similar to that of the Milky
Way.
 
We explore two questions in this paper (a) what is the shape of the
dark matter halo of this disk galaxy as a function of radius and what
types of orbits self-consistently produce this shape?  (b) is it
possible to use phase space coordinates to associate halo stars with
individual progenitor satellites in a fully cosmological context where
baryonic physics may have altered the dynamics non-adiabatically?
These are important questions in the context of modeling the Milky Way
with stellar orbits and will become increasing important in the next
few years with the advent of numerous resolved star surveys.  One of
the objectives of current and future surveys is to determine the global shape
and density profile of the Milky Way halo either by modeling the
kinematics of local field halo stars \citep{loebman_etal_12}, and/or
combining this with models of  individual tidal streams \citep[e.g.][]{johnston_etal_99, law_majewski_10}. 

The motivation for our investigation of the first question is that
although cosmological $N$-body simulations produce dark matter halos
that are triaxial or prolate \citep{jing_suto_02}, it has been known
for over a decade that the shapes of DM halos are altered when baryons
cool and condense at their centers. With baryons, halos become
spherical or axisymmetric within the inner one third of the virial
radius, but remain triaxial or prolate at larger radii
\citep{dubinski_carlberg_91, kazantzidis_etal_04shapes, deb_etal_08}.
If a gravitational potential is exactly oblate and axisymmetric, all
orbits in the system are axisymmetric short axis tubes
\citepalias{BT}. The prevailing view is that as a triaxial system
becomes more oblate due to chaotic mixing or other processes, an
increasing fraction of the box orbits and long axis tube orbits that
constitute the ``back bone'' of the triaxial system are converted to
axisymmetric short-axis tubes \citep{merritt_valluri_96}. Although it
is generally assumed that the self-consistent distribution function of
a nearly oblate system may be assumed to be comprised exclusively of
axisymmetric short-axis tubes, it has been shown that even small
deviations from axisymmetry can result in a large fraction of box
orbits in the inner regions of the potential
\citep{vandenbosch_dezeeuw_10}. Using controlled simulations in which
a rigid particle disk grows adiabatically inside a triaxial halo,
\citet{valluri_etal_10} (hereafter \citetalias{valluri_etal_10}) and \citet{valluri_etal_12} (hereafter \citetalias{valluri_etal_12}) 
showed that although the halos become more oblate in the inner
regions, a significant fraction of the orbit population retains its
original orbital characteristics (e.g. box orbits remain box orbits
and long axis tubes remain long-axis tubes). However, as the potential
changes shape adiabatically, the halo orbits with small pericenter
radii also become ``rounder'', but a large fraction do not change
their orbital type because their orbital integrals of motion (or
rather {\it orbital actions}) are adiabatic invariants. When a disk
galaxy is formed in more realistic hierarchical galaxy formation
simulations, the evolution is no longer adiabatic. Although several
recent studies \citep{tissera_etal_10, kazantzidis_etal_10,
  zemp_etal_12} find that the shapes of dark matter and stellar halos
in cosmological simulations are also nearly oblate, effects of
baryonic processes on the orbits of halo stars and dark matter
particles in Milky Way sized disks in this context is less well
understood. A recent study \citep{bryan_etal_12} suggests that orbital
properties of stars and dark matter particles in $10^{13}$\msun\ halos
may be somewhat sensitive to the feedback prescriptions
employed. Knowing how the shape of a cosmological halo varies with
radius and how DM and stellar orbits respond can provide vital
insights into the shape and formation history of our own halo.

The motivation for investigating the second question is to study the
impact of baryonic processes on ``Galactic Archeology'' with halo
stars. Although halo stars probably constitute no more than 1\% of the
stellar mass of the Galaxy, current and future surveys are focused on
uncovering their kinematical, spatial and abundance distributions.
This is because it has been argued that the time it takes for the
integrals of motion of the orbits of halo stars to change may be
longer than the age of the Galaxy, allowing correlations between the
abundances and kinematics of halo stars to serve as a ``fossil''
record of the formation history of the Galaxy. It has been shown that
deconstructing the accretion history of the halo (i.e. identifying
individual accretion events) is possible using orbital integrals of
motion in controlled simulations
\citep{helmi_dezeeuw_00,mcmillan_binney_08,gomez_etal_10}. However,
the effects on the orbital integrals of motion of hierarchical
evolution accompanied by dissipation, star formation and feedback from
baryons, have yet to be examined.

In addition, there has been significant recent effort devoted to interpreting the kinematic, metallicity and orbital distributions of halo stars to infer the formation history of the stellar halo. Theoretical studies have identified three possible origins for stars that currently reside in the halo: (a) they formed in a satellite galaxy outside the virial radius and were subsequently accreted  \citep[the so called ``accreted'' halo,][]{bullock_johnston_05}, (b) they formed inside a satellite after it is accreted by the main galaxy  \citep[``{\it in situ} halo stars'' or ``endo-debris''][]{tissera_etal_13}, (c)  they formed in the disk and were then kicked into the halo by infalling satellites  \citep[``kicked-out'' stars,][]{zolotov_etal_10, purcell_etal_10,sheffield_etal_12}.  Recent observations of halo stars (both local and distant field samples) provide evidence for the existence of at least two overlapping stellar halos with distinct metallicity and rotation signatures  \citep[e.g.][]{carollo_etal_07,carollo_etal_10, beers_etal_12, hattori_etal_13,kafle_etal_12}, however concerns have been raised about the distances measurements used in some of these analyses \citep{schoenrich_etal_11} and their dependence on the assumed rotation velocity of the local standard of rest \citep{deason_etal_11a}. We do not discuss the possible formation modes of halo stars in this paper, but refer readers to other recent works listed above. 

The outline of this paper is as follows: in Section~2 we describe the
simulations used in this study and the methods used to analyze the
orbits of stars and dark matter particles drawn from the simulations
\citep[Laskar frequency analysis][]{laskar_93, valluri_merritt_98,
  valluri_etal_10, valluri_etal_12}. In Section~\ref{sec:shape}, we
measure the shape of the dark matter (and stellar) halo as a function
of radius. We also analyze the types of halo orbits in both halos and
characterize their shapes and eccentricities as a function of
radius. In Section~\ref{sec:phasespace} we compare the phase-space 
distributions of halo stars and dark-matter particles using frequency maps. Finally in Section~\ref{sec:metals} we assess how baryonic
condensation alters the orbits of the tidal debris that constitutes
the stellar halo by examining the metallicity-formation time relation
for halo stars and their orbital properties.  In
Section~\ref{sec:discuss} we summarize our results, and discuss their
implications.

\section{Simulations and Numerical Methods}
\label{sec:method}

\subsection{Simulations}
\label{sec:simulations}

The MUGS sample of galaxies were produced using an $N$-body+SPH
simulation of the formation of galaxies in a fully cosmological
context.  Dark matter halos of mass $\sim10^{12}$~\msun\ at $z=0$ were
identified in a dark-matter-only ($N$-body) $\Lambda$CDM \citep[WMAP3
  parameters][]{spergel_etal_07} simulation within a box of comoving
length $50/h$~Mpc, without regard to halo spin or mass accretion
history.  These halos were then re-simulated using GASOLINE
\citep{gasoline} at high resolution using the volume renormalization
technique to allow for high resolution in the region of interest while
simultaneously including the full tidal field. The simulations include
metal cooling, heating from UV background radiation, star formation,
stellar energetic and metal feedback, and metal diffusion. Star
formation and stellar feedback are included via recipes based on the
``blastwave model'' \citep{stinson_etal_06, stinson_etal_10}. In the
high resolution region of these simulations, a dark matter particle
has a mass of $1.1 \times 10^6$~\msun, a gas particle has an initial
mass of $2.2\times 10^5$~\msun, and star particles form with masses of
$<5.5\ \times 10^4$~\msun. All particles have a gravitational
softening length of $\epsilon=312.5$~pc. Full details of the MUGS
simulations and properties of the sample galaxies can be found in
\citet{stinson_etal_10}.  The disk galaxy analyzed in this paper
(g15784) is the largest of the disk galaxies from the MUGS sample.

The galaxy g15784 contains a bulge which is fitted by a de Vaucouleurs
$R^{1/4}$ profile with an effective radius of $r_e=0.67$~kpc, and a
disk with an exponential scale length of 3.38~kpc. The maximum halo
circular velocity is 360~\kms and its virial radius ($r_{vir}$) is
290.8~kpc. We use the definition of $r_{vir}$ \citep{bryan_norman_98}
which corresponds approximately to the radius within which the average
density of the halo is $\sim 100$ times the critical
density.\footnote{Note that in V10, V12, $r_{200}$ was used.}  This
galaxy experienced its last major merger at $z=2.5$ about 10~Gyr prior
to the time at which we examine the simulation.

Cosmological simulations predict that the dark matter halos of all
galaxies contain hundreds to thousands of dark subhalos
\citep{moo_etal_98}, although only about 10\% of the total mass of the
halo is in such subhalos.  Subhalos in g15784 were identified by the
AMIGA halo finder AHF\footnote{http://popia.ft.uam.es/AMIGA/}
\citep{AMIGA_1_04,AMIGA_2_09}.  To determine the frozen potential in
which to integrate orbits of stars and dark matter particles in
g15784, we included mass within 4 virial radii to avoid an unphysical
discontinuous density change beyond the virial radius. This region
included several small satellite galaxies and numerous dark subhalos
and a small mass contribution at the edge of the simulation volume
from a neighboring $10^{12}$~\msun\ galaxy \citep{nickerson_etal_11}.
Both the central stellar spheroid and luminous subhalos are
unrealistically massive due to overcooling. Both dark and luminous
satellites can perturb orbits of halo particles.  Subhalos found by
AHF were removed; however, a few subhalos of mass $\sim
10^7$~\msun\ remained in the inner regions of the galaxy, where
automated halo finders have the most difficulty distinguishing
substructure \citep{knebe_etal_11}, making it difficult to fully
assess the effects of the subhalos on the orbits of stars and dark
matter particles.  We tested the characteristics of orbits integrated
both with and without the gravitational effect of subhalos included in
the frozen potential and found a higher fraction of chaotic orbits
when subhalos were included. We also carried out controlled
simulations where subhalos were allowed to evolve
(\citealt{deb_etal_08} --- hereafter \citetalias{deb_etal_08}). We
find both frozen and evolving subhalos increase the fraction of
chaotic orbits.  However the population that is most strongly affected
is the resonant orbit family. These orbit which are ``thin''
(i.e. they occupy only a small volume of physical space), were shown
to be important in halos where disks grow adiabatically (\citetalias{valluri_etal_12}). When subhalos are ``frozen'' in place,
a resonant orbit may repeatedly encounter a subhalo that lies close to
that volume causing it to become chaotic.  In this paper we only
consider orbits integrated in the frozen potential from which subhalos
more massive than $10^7$~\msun\ were removed. This does not reduce the
potential for scattering from smaller subhalos or irregularities such
as tidal streams however, and we will see that this scattering
probably contributes to the large fraction of chaotic orbits.

We compare the orbits of particles from the cosmological simulation to
orbits drawn from two controlled simulations of disk galaxies grown
adiabatically inside dark matter halos (from
\citetalias{valluri_etal_10} \&\citetalias{valluri_etal_12}).

In the controlled simulations the disks form quiescently and hence
alter the halos adiabatically. They also do not have subhalos and
other irregularities arising from hierarchical structure
formation. The first controlled simulation SA1 (see
\citetalias{deb_etal_08}, \citetalias{valluri_etal_10} \&
\citetalias{valluri_etal_12} for details) is a triaxial dark matter
halo in which a disk of particles representing a baryonic mass
fraction of 2.5\% was grown (starting from nearly zero initial mass)
adiabatically and linearly on a timescale $t_g=5$~Gyr with the disk's
symmetry axis parallel to the short-axis of the halo.  The triaxial
$N$-body halo used in simulation SA1 was produced by multi-stage
merger of spherical NFW halos.  In the second controlled simulation,
SBgs, a baryonic stellar disk forms self-consistently from $2.8
\times10^{11}$~\msun\ of initially hot gas (constituting 10\% of the
total mass) in a prolate NFW halo (\citetalias{valluri_etal_12}, \citet{debattista_etal_13}). In SBgs the hot halo gas initially
has the same spatial distribution as the dark matter and slowly cools
to form a disk with symmetry axis along the short axis of the halo.
The halo and gas particles are given an initial specific angular
momentum $j$, determined by overall cosmological spin parameter
$\lambda = (j/G)(|E|/M^3)^{1/2} = 0.039$, which is motivated by
cosmological $N$-body experiments \citep{bullock_etal_01}.  The
simulation closely follows that described in \citet{roskar_etal_08}
and is evolved with the parallel $N$-body+SPH code GASOLINE
\citep{gasoline} for 10 Gyr.  Table~\ref{tab:simulations} summarizes
some of the key properties of the different simulations used in this
paper.

\begin{table*}
\caption{Details of simulations}
%\begin{centering}
\begin{tabular}{llllllll}\hline \\
%\begin{centering}
\multicolumn{1}{c}{Run } &
\multicolumn{1}{c}{$r_{vir}$} &
\multicolumn{1}{c}{$M_{vir}$\footnote{$M_{vir}$: mass within the virial radius ($r_{vir}$).}}&
\multicolumn{1}{c}{$M_b$\footnote{$M_b$: mass in baryons.}}  &
\multicolumn{1}{c}{$\epsilon_{DM}$\footnote{$\epsilon_{DM}$: gravitational softening length}} &
\multicolumn{1}{c}{$h$ \footnote{$h$: exponential radial scale length of disk.} }&
\multicolumn{1}{l}{Run Description }&
\multicolumn{1}{l}{Reference \footnote{Reference where the simulation is described.}} \\ 
 Name   &  [kpc] & [$10^{12}~\msun$] & [$10^{11}~\msun$]& [pc]            &  [kpc]  &                   &         \\ \hline\\
g15784 &  290.8 & 1.4                    & 2.1                       & 312.5        & 3.38 &{MUGS disk galaxy}                        &  \citet{stinson_etal_10}\\
SA1      &   491.9  &  6.9                    & 1.75                     & 100           & 3.0   &{Triaxial halo+short-axis stellar disk} & D08, \citetalias{valluri_etal_12} \\ 
% 661: prolate halo + short-axis starforming gas disk:
SBgs     &  317.9   &  1.9                 &         0.9  \footnote{Mass of baryons in stars only.}        & 100          &  1.9     &{Prolate halo+hot gas $\rightarrow$ disk w/ SF}   &  Debattista et al. (in prep)\\
\hline
\end{tabular}
 \label{tab:simulations}
\end{table*}

\subsection{Halo Orbit Sample}

In each simulation we selected $1-2\times10^4$ halo particles,
randomly distributed within some volume. The ``global sample'' refers
to particles selected at random within a spherical radius $r_g$
centered on the model's galactic center (potential minimum).

In the MUGS galaxy g15784 we identify halo stars by excluding stellar
orbits associated with the massive central spheroid (orbits with
apocenter radius $\rapo <5$~kpc) and stellar disk (orbits whose
maximum distance from the disk plane $\zmax < 3$~kpc ).  The global
sample comprised $\sim 10^4$ halo stars with $r_g <50$~kpc.  Since the
controlled simulations do not form stellar halos, we follow the orbits
of dark matter particles selected with $r_g<200$~kpc in these two simulations.

In all three potentials orbits were integrated forward from their
$z=0$ initial conditions for a duration of $\sim 50$~Gyr in a frozen
potential resulting from the full mass distribution of the simulation
including dark matter and baryons using an integration scheme based on
the {\sc PKDGRAV} tree. The assumption made here (and by most studies
that involve the kinematics of halo stars in Milky Way) is that the
potential of the galaxy at $z=0$ is nearly in steady state.  While
this assumption is not strictly true, there is evidence to support the
view that halos of mass similar to the Milky Way, evolve less at the
present epoch than they have in the past. \citet{busha_etal_07} show
that halos with $M_{200} \lesssim 10^{13}$~\msun\ in $\Lambda$CDM
simulations evolved for 64~$h^{-1}$~Gyr after the Big Bang (see their
fig 8) grow rapidly till $t/h = 10$ (corresponding to 13.8~Gyr after
the Big Bang) after which there is little change in their virial
mass. This does not mean that secular evolution in the disk and halo
stops at $z=0$, merely that rate of mass accretion slows down. We
emphasize that the long orbital integration times only serve to
improve the accuracy with which orbits in the potential at $z=0$ can
be characterized. We do not imply that the potential will have been
static for the duration of integration nor imply that the long
integration time is physically meaningful.

\subsection{Laskar Frequency Analysis}
\label{sec:laskar}

In this section we provide a brief description of the Laskar's
frequency analysis method. A more detailed description of the
application to orbits in $N$-body simulations is given in
\citetalias{valluri_etal_10} and \citetalias{valluri_etal_12}.

Orbits in galaxies are approximately quasi-periodic 
(\citealp{BT}; hereafter \citepalias{BT}),
hence their space and velocity coordinates can be
represented by a time series of the form: $x(t) = \sum_{k=1}^{k_{max}}
A_{k} e^{i\omega_kt}$, and similarly for $y(t), v_x(t)$ etc. This
means that a Fourier transform of such a time series will yield the
spectrum of frequencies $\omega_k$, and amplitudes $A_k$, that define
the orbit \citep{binney_spergel_82, binney_spergel_84}. In a three
dimension potential, only three frequencies in the spectrum are
linearly independent. The remaining $\omega_k$ are given by $\omega_k
= l_k\Omega_1+m_k\Omega_2+n_k\Omega_3$. For a regular orbit the
frequencies $\Omega_i, i=1...3$ are constant ``fundamental
frequencies'' that are related to 3 angle variables $\theta_i(t) =
\Omega_i t$ and 3 integrals of motion (more precisely actions $J_i$)
\citep{BT}. Since the full phase-space trajectory in the action-angle
coordinates $(J_i,\theta_i)$ resembles a 3D torus, the mapping from
$(\mathbf {x}(t), \mathbf{v}(t)) \Leftrightarrow ({\mathbf J,\theta})$
is referred to as {\it torus construction}.  \citet{laskar_90,
  laskar_93} developed a very accurate numerical technique ``Numerical
Analysis of Fundamental Frequencies'' to recover frequencies by taking
Fourier Transforms of complex time series of the form $\mathbf
{x}(t)+i \mathbf{v}(t)$. We use the implementation of
\citet{valluri_merritt_98}, which uses integer programming to recover
orbital fundamental frequencies with extremely high accuracy in $\sim
20-30$ orbital periods (significantly faster than 100s of orbital
periods needed by other codes; e.g., \citealp{carpintero_aguilar_98}.
Orbits were integrated and orbital frequencies computed in Cartesian
coordinates and in cylindrical coordinates, as described in
\citetalias{valluri_etal_12}.

Orbital fundamental frequencies can be used to: (i) classify orbits
into major orbit families \citep{carpintero_aguilar_98}; (ii) quantify
the {\it elongation} of an orbit relative to the shape of the
potential (\citetalias{valluri_etal_10}); (iii) identify chaotic
orbits \citep{laskar_93}; (iv) identify important resonant orbit
families \citep{robutel_laskar_01} which have trapped a large number
of resonant orbits yielding insights into the importance of secular or
adiabatic processes; and (v) determine the self-consistent equilibrium
potential from which a population of orbits was drawn
(\citetalias{valluri_etal_12}).

To determine the fraction of chaotic orbits in a potential, the
orbital time series is divided into two equal segments and the orbital
fundamental frequencies computed in each time segment. Since regular
orbits have fixed frequencies that do not change in time, the change
in the frequency measured in the two time segments can be used to
measure the drift in frequency space. \citetalias{valluri_etal_10}
showed that even for orbits in $N$-body potentials (which are
inherently noisy), it is possible to distinguish between $N$-body
jitter and true chaos by a quantitative measurement of frequency drift
by defining a diffusion rate parameter $\log(\Delta f)$ which measures
the logarithm of the change in the frequency of the leading term in
the orbit's frequency spectrum, measured in two consecutive time
segments.  \citetalias{valluri_etal_10} showed, using orbits in
$N$-body simulations of spherical NFW halos with gravitational
softening $\epsilon=100$~pc, that orbits with $\log(\Delta f)< -1.2$
were regular. In g15784 the fraction of orbits with $\log(\Delta f)>
-1.2$ (chaotic by the criterion established in the spherical NFW halo)
is over 50\%. However visually many of the orbits have almost regular
appearances. g15784 has a larger gravitational softening
($\epsilon=312.5$~pc) which appears to result in a slightly smoother
potential and hence orbits with $\log(\Delta f)< -0.5$ appear more
``regular'' (i.e. they fill a well defined volume in configuration
space) than they do in the controlled simulations. We therefore use
this more relaxed criterion for identifying regular orbits in the
cosmological simulation, {\it but it must be kept in mind that in
  $N$-body simulations all orbits are inherently ``noisy'' resulting
  in a continuous distribution of diffusion parameters $\log(\Delta
  f)$ and any cut-off between ``regular'' and ``chaotic'' is
  arbitrary}. Furthermore we have found that decreasing the cutoff to
$\log(\Delta f)< -1.2$ does not significantly affect the fractions of
short- and long-axis tubes, but decreases the fraction of box orbits
while increasing the fraction of chaotic orbits. We therefore caution
readers that in this cosmological simulation the boundary between chaotic and regular box orbits is
rather fuzzy and although we distinguish between them for consistency
with our previous work, they are better thought of as a single family
of ``centrophilic'' orbits with no net angular momentum.

\section{Results}
\label{sec:results}

In this section we examine various properties of the stellar and dark matter halos of the simulated disk galaxy g15784.  In particular we examine the shape of the dark matter and stellar halos (Section ~\ref{sec:haloshape}) the orbital properties of the stars and dark matter particles  (Section~\ref{sec:orbtyp}), and the phase space distributions of halo stars and dark matter particles (Section~\ref{sec:phasespace}). It is important to keep in mind that although we expect that many of these results are likely to be typical of dark matter and stellar halos from cosmological hydrodynamical simulations, we are in fact, analyzing only one halo in a cosmological context and so the trends and conclusion found should be taken with caution since they might depend on the accretion history of the particular halo and on the sub-grid physics. 

\subsection{Halo shape and orbital properties}
\label{sec:shape}

\subsubsection{Halo shape}
\label{sec:haloshape}

\begin{figure}
%\centering \includegraphics[trim=0.pt 0.pt 0.pt 0.pt, width=0.4\textwidth]{g15784_shape_lin_halo.ps}
\centering \includegraphics[trim=0.pt 0.pt 0.pt 0.pt, width=0.4\textwidth]{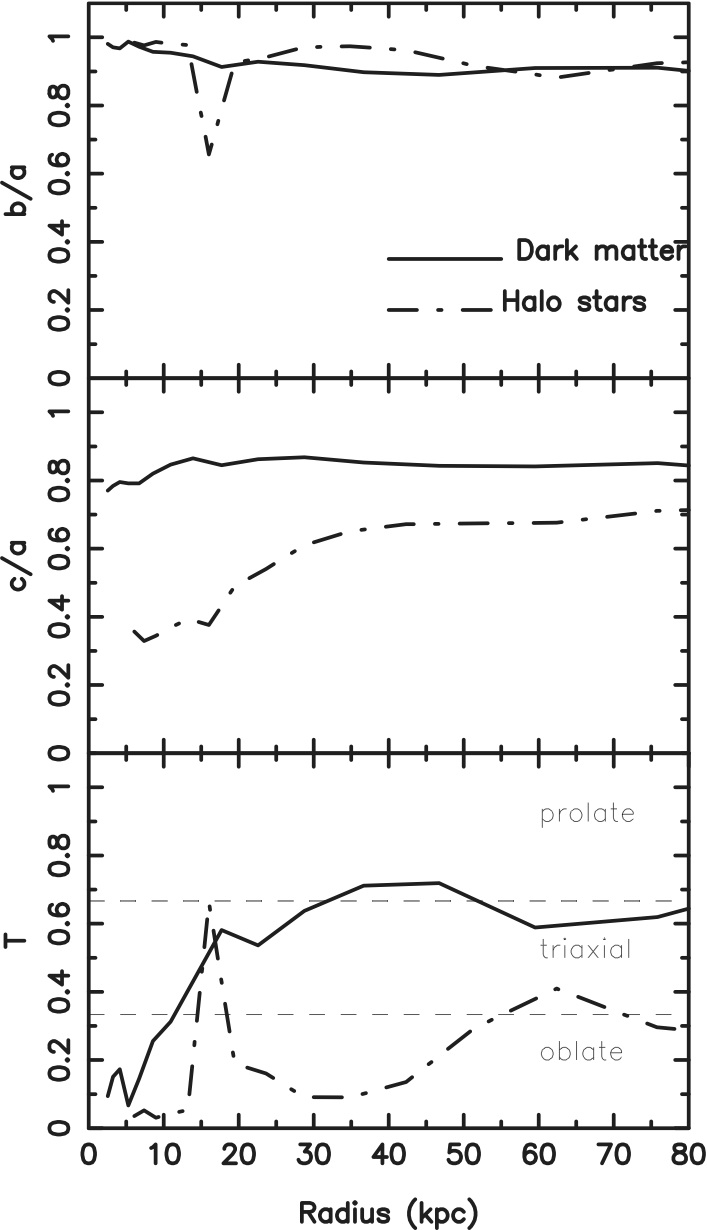}
\caption{Axis ratios $b/a$ (top) and $c/a$ (middle), triaxiality $T$ (bottom) versus radius from the center of galaxy g15784 for: dark matter particles (solid lines)
  and halo stars (dot-dashed lines). The dark matter halo is oblate in the inner $\sim 15$~kpc  transitioning to triaxial ($0.33 <T < 0.66$) at larger radii. The stellar halo is nearly oblate within  within $\sim 50$~kpc ($T \sim 0.2$) and becomes slightly triaxial beyond this radius.}
\label{fig:shape}
\end{figure}

We computed the shapes of the spatial matter distributions of dark
matter and star particles in g15784 within ellipsoidal shells centered
on the minimum of the potential. The shape is determined at each
radius $r$ by finding the ellipsoidal shell of width $\sim 5$~kpc
whose principal axes $\mathbf{a}$, $\mathbf{b}$, and $\mathbf{c}$
self-consistently (1) have a geometric mean radius $\sqrt{abc}=r$, and
(2) are the eigenvectors of the second moment tensor of the mass
distribution within the ellipsoidal shell (J. Bailin et al. 2013, in
preparation). In practice, this is done by starting with a spherical
shell, calculating the second moment tensor, deforming the shape of
the shell to match the eigenvectors of the tensor, and iterating until
the solution has converged; this is very similar to the method
advocated by \citet{zemp_etal_11}.

To measure the shape of the stellar halo (at $z=0$) we select halo
stars by excluding all stars within 5~kpc of the center of the galaxy
(defined as the ``bulge''), as well as all stars which lie within
3~kpc of the disk plane (defined as a thin+thick disk) for particles
within 20~kpc (the radius within which most disk particles
reside). This cut to identify ``halo stars'' is not perfect since it
incorrectly excludes the small fraction of halo stars that happen to
lie close to the disk plane or within 5~kpc of the galactic center at
$z=0$.  This cut may potentially bias $c/a$ values to be somewhat
higher than they should be within the 20~kpc inner region of the halo,
but is unlikely to significantly alter $b/a$. We use a more
sophisticated method (relying on kinematic properties) to exclude disk
stars when we examine orbital properties of stars.

\begin{figure*}
\centering \includegraphics[trim=0.pt 0.pt 0.pt 0pt,
 width=0.34\textwidth]{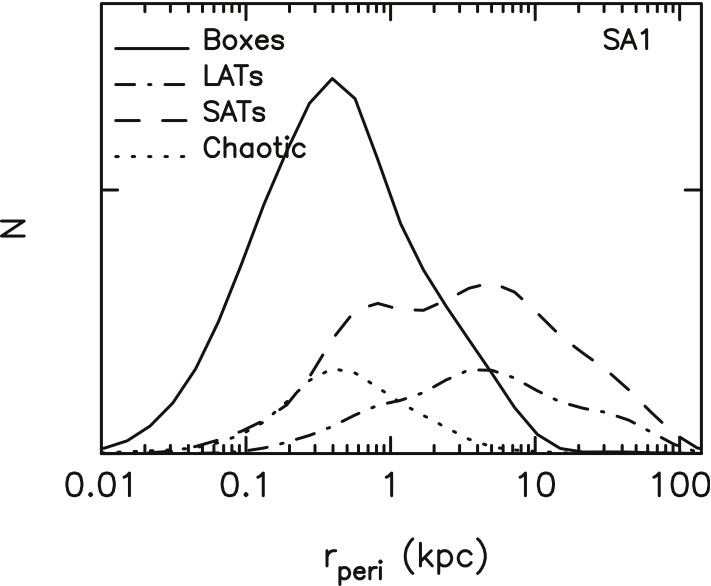}
\centering \includegraphics[trim=0.pt 0.pt 0.pt 0.pt,
 width=0.34\textwidth]{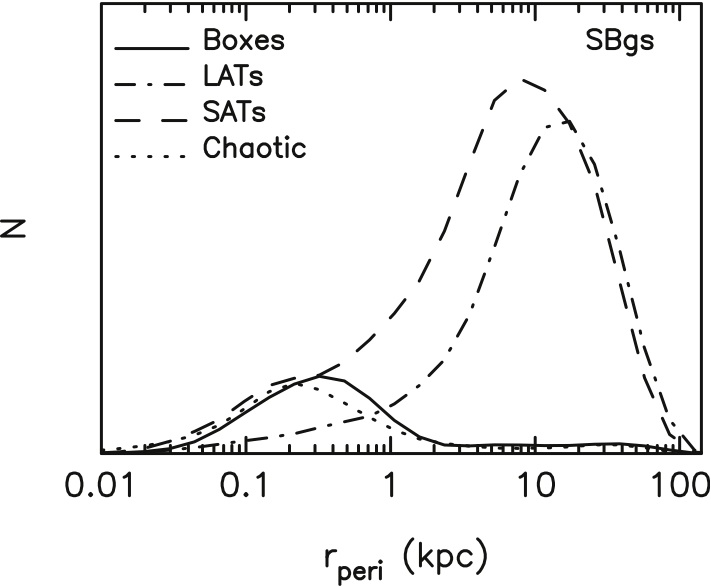}
\centering \includegraphics[trim=0.pt 0.pt 0.pt 0pt,
 width=0.34\textwidth]{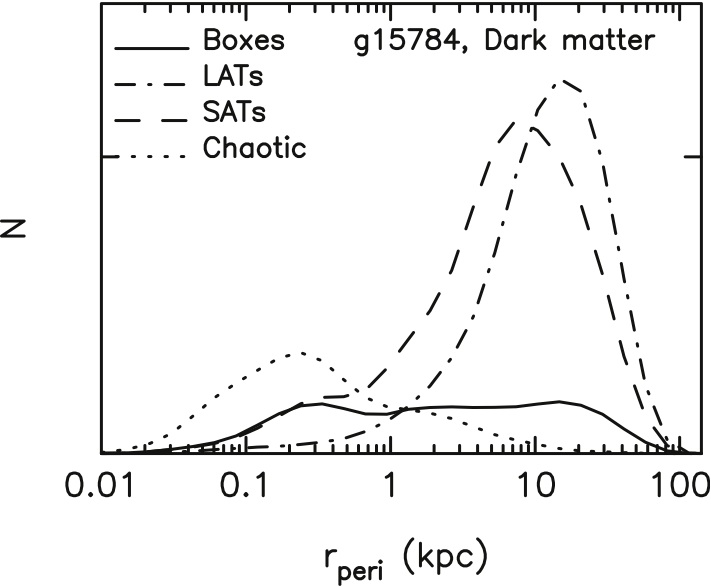}  
\centering \includegraphics[trim=0.pt 0.pt 0.pt 0pt,
width=0.34\textwidth]{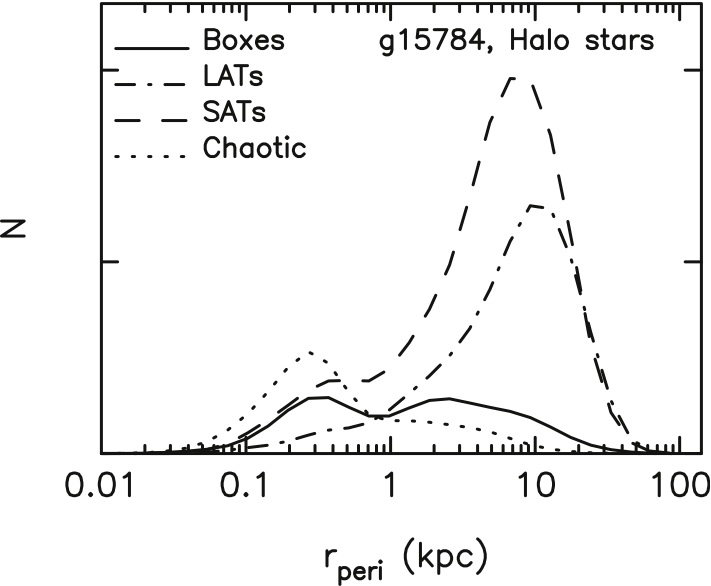}%chaolim=-0.5
% boxes, xtube, ztub, chaotic
 %       1063        2616        4333        1128
 % 0.11630197      0.28621444      0.47407001      0.12341357   
  \caption{Kernel density distributions of orbit types as a function
    of orbital pericenter distance $\rperi$ for $10^4$ halo particles
    in 3 simulations. In each panel the vertical axis is proportional
    to the number of orbits with a particular value of $\rperi$, i.e.
    the sum of integrals over all curves in a panel is equal to the
    total number of orbits. Top left: orbits of dark matter particles
    from adiabatic $N$-body simulation SA1 (triaxial dark matter halo
    with stellar disk). Top right: orbits of dark matter particles
    from adiabatic simulation SBgs (prolate dark matter halo with
    stellar disk formed from hot gas).  Bottom left: dark matter
    particles in cosmological simulation g15784.
    Bottom right: halos stars in g15784. Line
    styles indicate the 4 major orbit families.  Only orbits with
    periods shorter than 2.5~Gyrs were selected. In the top two plots
    the spatial gravitational softening length is 0.103~kpc, while in
    the bottom two panels it is 0.3125~kpc. The orbital distributions below the softening radius in each model are unreliable.}
\label{fig:orbtyp_peri_r50}
\end{figure*}

Following standard nomenclature $a, b, c$ are defined as the semi-axis
lengths of long ($x$), intermediate ($y$) and short ($z$) axes
respectively. The short axis is perpendicular to the plane of the disk
in the inner region (the disk warps beyond about 30~kpc).
Figure~\ref{fig:shape} shows axis ratios as a function of radius $b/a$
(top panel) and $c/a$ (middle panel), and the triaxiality parameter $T= (1 -b^2/a^2)/(1-c^2/a^2)$ (bottom panel).  The triaxiality parameter $T$ is often used to characterize the shape of ellipsoidal figures which are termed ``oblate'' if $T<1/3$, ``triaxial'' if $1/3< T< 2/3$ and ``prolate'' if $T>2/3$. Thin dashed horizontal lines marking these shape regimes are marked in the bottom panel of Fig~\ref{fig:shape}. 

For the dark matter halo, $b/a$ and $c/a$ vary only slightly with radius ($1< b/a < 0.9$, $c/a \sim 0.8$). The bottom plot shows that the halo is  oblate within $\sim 15$~kpc, becoming increasingly triaxial  to prolate-triaxial beyond this radius. The shape of the stellar halo is highly oblate ($T \sim 0.1$)  within 40~kpc, becoming mildly triaxial at larger radii. The stellar halo is also much flatter ($c/a$ is smaller)  than the dark matter halo at all radii. Note that at radius $\sim 17$~kpc, the sharp dip in $b/a$  and the corresponding spike in $T$ are the result of a subhalo of mass $\lesssim 10^7$~\msun\, that was not identified by the Amiga halo-finder. \citet{zemp_etal_11} show that when subhalos are not subtracted (or improperly subtracted) they produce spurious spikes in the shape parameters. The overall changes in the shape parameters (e.g. $T$) of the dark matter halo and stellar halo as a function of radius are consistent with the results of previous studies \citep[e.g.][]{zemp_etal_12}.

\subsubsection{Orbital type distributions}
\label{sec:orbtyp}

To assess how orbital structure depends on shape, we examined orbits
of $\sim 10^4$ star and dark matter particles in the global sample of
the galaxy g15784, and for comparison, $10^4$ dark matter orbits from
each of the two controlled simulations SA1 and SBgs. Orbits were
classified into four types (box orbits, short-axis tubes [SAT], long
axis tubes [LAT], and chaotic orbits). Their distributions as a
function of orbital pericenter radius ($\rperi$) and orbital
eccentricity were examined in order to understand how the shapes of
the two components (dark matter and halo stars) are related
to the types of orbits that self-consistently reproduce their shapes.

Figure~\ref{fig:orbtyp_peri_r50} shows orbit populations as a function
of $\rperi$ in the two controlled simulations SA1 and SBgs (top
panels), and in g15784 (bottom panels). We choose to plot the orbit
fractions as a function of $\rperi$ rather than some other measure of
the ``radius'' of an orbit (such as the time-averaged radius) because
we are specifically interested in understanding if the SATs contribute
to the oblateness in the inner halo. Since $\rperi$ defines the inner
boundary of a tube orbit, it defines the region interior to which the
orbit contributes no mass. Hence this quantity is best suited to
addressing this question.  In theory, box orbits and chaotic orbits
can travel arbitrarily close to the center of the potential and so
formally they have $\rperi=0$ but, in practice $\rperi>0$ during the
50~Gyr (and at least 20 orbital period) integration time.  Each panel
shows kernel density distributions of each of the four orbit families
as indicated by the line legends.  The area under each curve is
proportional to the total number of orbits of that family. It is
important to note that in the top two panels the gravitational
softening length of dark matter particles is $\sim 0.1$~kpc, while in
the bottom panels the gravitational softening length is $\sim
0.3$~kpc; distributions at pericenter radii smaller than the softening
lengths are not reliable.
 
The progenitor halo of SA1 was triaxial with 86\% of particles on box
orbits and 11\% on LATs (\citetalias{valluri_etal_10}).
Figure~\ref{fig:orbtyp_peri_r50} shows that this population was
transformed (due to the growth a stellar disk) to a halo that is still
dominated by boxes (49\%) but the fraction of SATs (32\%), LATs
(12\%), and chaotic orbits (7\%) has increased significantly. In
contrast the progenitor halo of SBgs was prolate with 78\% LATs and
15\% box orbits (\citetalias{valluri_etal_10}). Following the
condensation of hot gas and the growth of a stellar disk, the dark
matter halo orbit population evolves to an overall potential in which
SATs dominate (51\%) with LATs (35\%) also important. In the
cosmological simulation g15784, halo star particles
(Fig.~\ref{fig:orbtyp_peri_r50}~bottom right) are primarily on SATs
(47\%) and LATs (29\%) and dark matter particles also have similar
orbit fractions (40\% are on SATs and 36\% are on LATs).

Table~\ref{tab:orb_frac} shows the orbit fractions in the inner and
outer regions of all three halos following the growth of disks. We
computed the relative fractions of different types of orbits with
$\rperi$ less than or greater than 20~kpc. Since the inner $\sim
$~20~kpc region is where the disk dominates, this is the region where
the shape of the halo is close to oblate axisymmetric in all three
potentials. In none of the halos do SATs constitute more than 50\% of
the orbit fraction in the oblate inner halo ($\rperi <20$~kpc). In all
four models the ``triaxial orbit families'' (box orbits and chaotic
orbits and long-axis tubes), together constitute 50\% or more of the
population. In SBgs, g15784, the outer part of the halo ($\rperi
>20$~kpc) are mildly triaxial. Here we find that LATs are the dominant
population. In SA1 -- which is  quite strongly triaxial beyond
20~kpc -- SATs constitute 65\% of the population with LATs constituting 33\%.
{\em Thus in no case   do we find evidence that an oblate density distribution implies a
  distribution function dominated by axisymmetric SATs, and conversely we find that the presence of a large fraction of short-axis tubes does not imply axisymmetry}.

Although the distributions of orbit types as a function of $\rperi$
differ in detail for dark matter particles and halo stars in g15784,
their overall distributions are significantly more similar to each
other than they are, for instance, to the orbit distribution of dark
matter particles in model SA1 (Fig.~\ref{fig:orbtyp_peri_r50}~top
left).  Furthermore g15784 shows a striking resemblance to the orbit
populations in SBgs. The similarity is probably coincidental in view
of the fact that g15784 formed in a cosmological context and has
experienced significant perturbations due to mergers and
substructure. Minor mergers are entirely absent from the history of
SBgs, which formed from a single major merger between two spherical
halos. We expect the halos of galaxies of the mass of g15784 to have
experienced multiple major mergers and to therefore be more like SA1,
rather than SBgs which has only had one major merger. It appears that
since g15784 had its last major merger nearly 10~Gyr ago, this
galaxy's stellar halo was primarily formed via minor majors (rather
than major mergers). Since the orbits at large radii are primarily
tubes, we hypothesize that the satellites that formed the stellar halo
were mainly accreted on tube orbits. We will see in
Section~\ref{sec:metals} that there is support for this hypothesis.

\begin{table}
\caption{Orbit fractions in inner and outer halo regions}
\begin{centering}
\begin{tabular}{l|cccc|cccc}\hline \\
%\begin{centering}
\multicolumn{1}{l|}{Run} &
\multicolumn{4}{c|}{$\rperi\leq 20$kpc } &
\multicolumn{4}{c}{$\rperi>20$kpc} \\
\multicolumn{1}{l|}{    } &
\multicolumn{1}{c}{Box} &
\multicolumn{1}{c}{Chaotic} &
\multicolumn{1}{c}{LAT} &
\multicolumn{1}{c|}{SAT}&
\multicolumn{1}{c}{Box} &
\multicolumn{1}{c}{Chaotic} &
\multicolumn{1}{c}{LAT} &
\multicolumn{1}{c}{SAT}\\ 
\hline\\
%       & B     &   C    &L (x) & S(z) &  B     &   C    & L (x) & S(z)  \\
SA1   & 0.51 & 0.09 & 0.11 & 0.29 & 0.01 & 0.05 & 0.33 & 0.65 \\
SBgs & 0.10 & 0.10 & 0.30 & 0.50 & 0.03 & 0.02 & 0.52 & 0.43 \\
Stars & 0.12 & 0.13 & 0.32 & 0.43 & 0.03 & 0.02 & 0.55 & 0.41 \\
DM    & 0.12  & 0.12 & 0.34 & 0.42 & 0.09 & 0.00 & 0.59 & 0.32 \\
\hline
\end{tabular}
\end{centering}
\label{tab:orb_frac}
\end{table}

Recently \citet{bryan_etal_12} analyzed orbits in halos of mean mass
$\sim 6\times 10^{13}$~\msun\ (at $z=0$) and $7\times
10^{11}$~\msun\ (at $z=2$) from the OverWhelmingly Large Simulations
(OWLS) to investigate the effects of various feedback prescriptions on
the orbital properties of stars and dark matter particles in a
cosmological context.  \citet{bryan_etal_12} selected 50 halos from 5
different simulations (with different feedback prescriptions) and
selected 500 particles per halo. They integrated orbits in
gravitational potentials computed with the Self-Consistent Field (SCF)
method and classified them using the method of
\citet{carpintero_aguilar_98}.

\citet{bryan_etal_12} find that as the central baryonic fraction
increases, the fraction of box orbits in the inner part of the halo
decreases. Since we only examine one simulation we are unable to
verify their result that the fraction of box orbits in the inner
region decreases with increasing central baryonic mass fraction;
however, the baryonic mass fraction in the inner 20~kpc region of
g15784 (where the disk dominates) is obviously much higher than in the
outer region, and we find no evidence that box orbits have been
depleted in the inner regions as a consequence of baryonic
condensation.  The difference between our result and those of
\citet{bryan_etal_12} is most probably a consequence of differences in
the way in which the total galactic potential (in which orbits are
evolved) is computed. The SCF method \citep[used by][]{bryan_etal_12}
uses a low-order multipole expansion code to compute the potential,
which consequently is much smoother and has more symmetry than our
potential, which is computed with the PKDGRAV tree-code and therefore
includes all the irregularities of the full cosmological simulation
(except subhalos more massive than $10^7$~\msun\ which were removed).

\subsubsection{Orbital shapes: elongation versus eccentricity}
%\label{sec:orbtyp}

\begin{figure*}
\centering \includegraphics[trim=0.pt 0.pt 0.pt 0pt,width=0.34\textwidth]{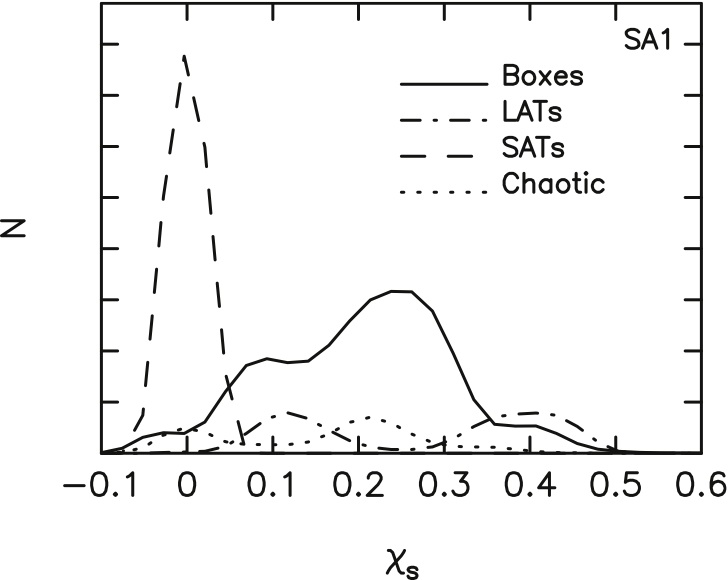}
\centering \includegraphics[trim=0.pt 0.pt 0.pt 0.pt, width=0.34\textwidth]{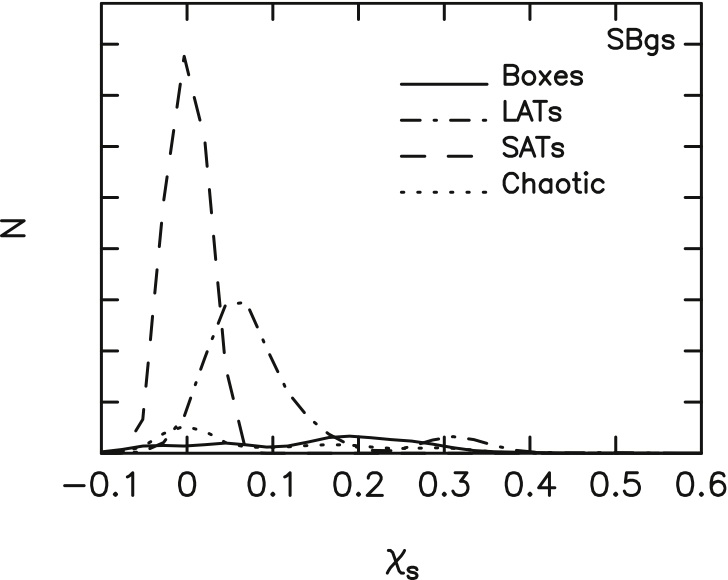}
\centering \includegraphics[trim=0.pt 0.pt 0.pt 0pt, width=0.34\textwidth]{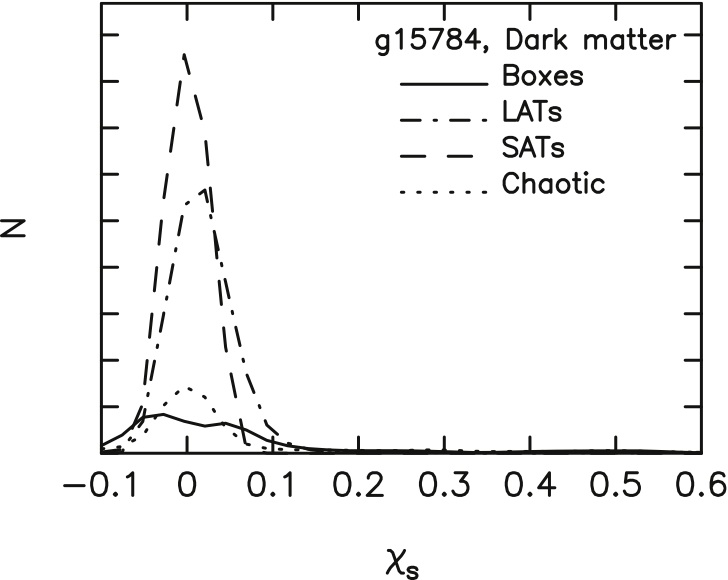}
% Chaotic limit -0.5 , all DM
\centering \includegraphics[trim=0.pt 0.pt 0.pt 0pt,width=0.34\textwidth]{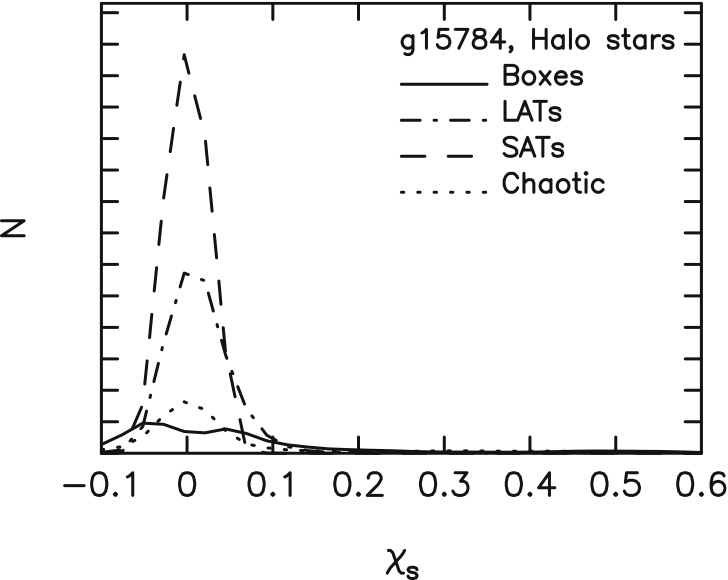}
% Chaotic limit -0.5 , halo stars
%boxes, xtube, ztub, chaotic
%        1063        2616        4333        1128
%  0.11630197      0.28621444      0.47407001      0.12341357   
\caption{Similar to Fig~\ref{fig:orbtyp_peri_r50}: Kernel density distribution of orbit types as a function of orbital elongation parameter $\chi_s$. Values of $\chi_s >0.1$ denote orbits elongated along the major ($x$) axis of the halo, while orbits with $\chi_s \sim 0$ are axisymmetric or ``round''.}
\label{fig:hist_chis}
\end{figure*}

In any self-consistent potential, the shapes of the majority of the
orbits must match the overall shape of the 3-dimensional matter
density distribution. In a triaxial potential, the elongation along
the major axis is provided either by box orbits or inner LATs.  The
ratios of the fundamental frequencies of orbits can be used to
characterize their overall shape. For a triaxial potential, the fact
that semi-axes lengths $a > b > c$ implies that the oscillation
frequencies obey the condition $|\Omega_x| < |\Omega_y| < |\Omega_z|$
for any (non resonant) orbit with the same over-all shape as the
density distribution (we consider only the absolute values of the
frequencies since their signs only signify the sense of oscillation).
\citetalias{valluri_etal_10} use this property to define an average
``elongation'' parameter $\chi_s$ for any orbit.  When an orbit is
elongated in the same way as the potential,
 \begin{eqnarray}
  & & |\Omega_z| >   |\Omega_y| >   |\Omega_x| \nonumber \\
  &\Rightarrow&  {\frac{|\Omega_y|}{|\Omega_z|}}   > {\frac{|\Omega_x|}{|\Omega_z|}}.\nonumber 
 \end{eqnarray}

Therefore we can define an elongation parameter $\chi_s$  by
 \begin{eqnarray}
   \chi_s  &\equiv& {\frac{|\Omega_y|-|\Omega_x|}{|\Omega_z|}},
   \end{eqnarray}
such that $\chi_s >0$ for orbits elongated along the major axis with
larger values of $\chi_s$ implying a greater the degree of elongation
(V10).  \citetalias{valluri_etal_12} showed that some systems can be
elongated along the $y$-axis at small radii but elongated along the
$x$-axis at larger radii; also some outer LAT orbits can have greater
extent along the $y$-axis than along the $x$-axis. In both these cases
$\chi_s$ is slightly negative. Orbits for which all frequencies are
almost equal enclose a volume that is almost spherical have $\chi_s
\sim 0$ because $\Omega_x \sim \Omega_y \sim \Omega_z$.  Also orbits that are close to axisymmetric about the short ($z$) axis (i.e. SATs) have  $\Omega_x \sim \Omega_y$ and hence  $\chi_s \sim 0$, regardless of the value of $\Omega_z$.
\begin{figure*}
\centering \includegraphics[trim=0.pt 0.pt 0.pt 0pt, width=0.34\textwidth]{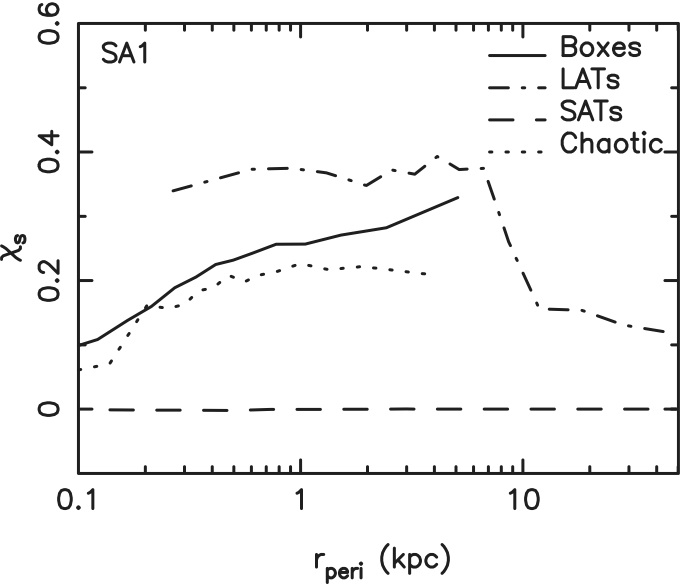}
\centering \includegraphics[trim=0.pt 0.pt 0.pt 0.pt, width=0.34\textwidth]{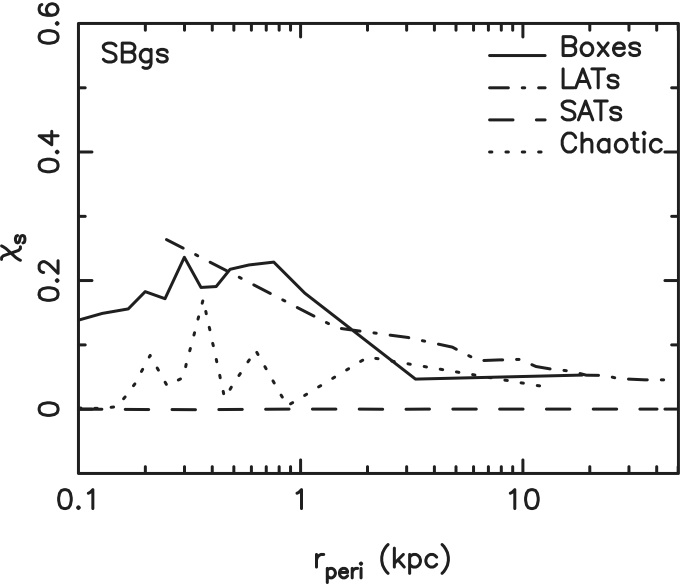}
\centering \includegraphics[trim=0.pt 0.pt 0.pt 0pt, width=0.34\textwidth]{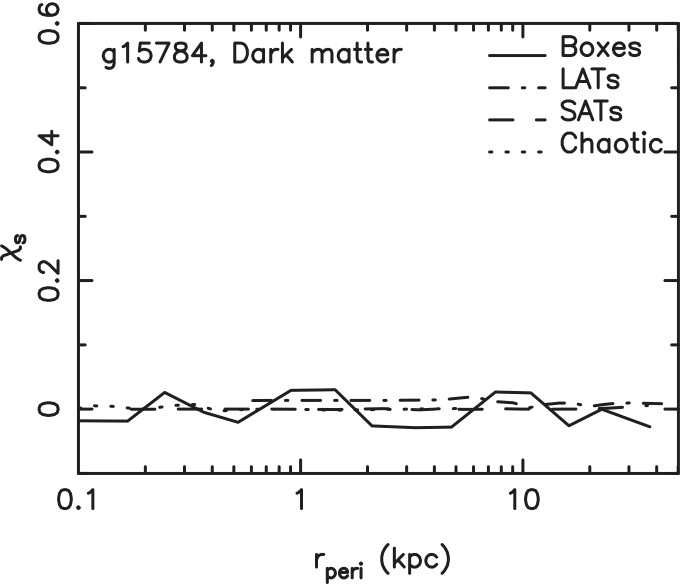}
% Chaotic limit -0.5 , all DM
\centering \includegraphics[trim=0.pt 0.pt 0.pt 0pt, width=0.34\textwidth]{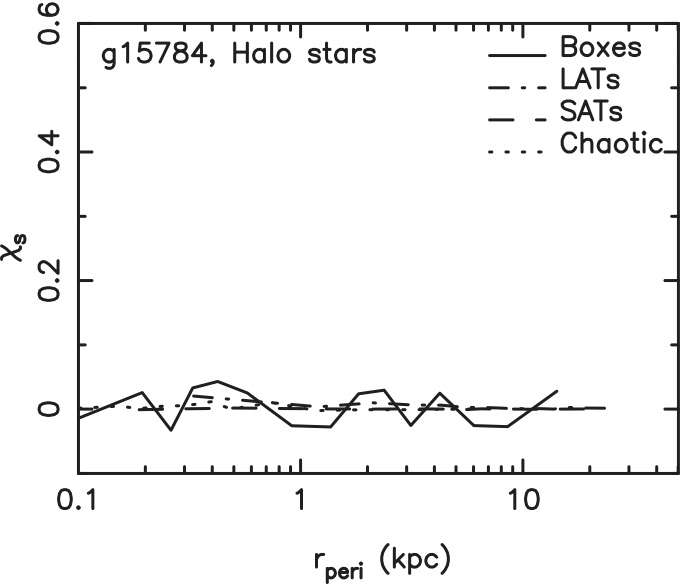}
% Chaotic limit -0.5 , halo stars
%boxes, xtube, ztub, chaotic
%        1063        2616        4333        1128
%  0.11630197      0.28621444      0.47407001      0.12341357   
\caption{The median value of the orbital elongation parameter $\chi_s$  in 15 bins in  $\rperi$ for orbits in each of the three models. }
\label{fig:orbtyp_sh_rp}
\end{figure*}

Figure~\ref{fig:hist_chis} shows kernel density distribution of orbits
of the four different types as a function of orbital elongation
$\chi_s$. SA1 shows significantly larger fraction of elongated orbits
($\chi_s >0.1$) than other models.  The SATs (dashed lines) have
$\chi_s = 0 $ because they are close to axisymmetric.  Model SBgs
(which is prolate at large radii) has a significant fraction of
elongated LAT orbits.  In contrast, for g15784, all four families of
orbits for both dark matter (bottom left) and stars (bottom right)
have very small elongation (i.e. they are quite ``round'' or
axisymmetric).

\citetalias{valluri_etal_10} showed for several controlled simulations
(including SA1) that the orbital elongation parameter $\chi_s$ in the
inner regions decreased from $\chi_s \sim 0.4 - 0.5$ in the triaxial
or prolate models to $\chi_s \sim 0.2$ after baryonic components were
grown. They showed that although some orbits do change from boxes or
LATs in the prolate systems to SATs, all orbits adapt to the more
axisymmetric baryonic component by becoming less elongated, especially
at small $r_{\rm peri}$. However, only a small fraction of all orbits
(between 4\% and 25\%) actually transform themselves to become SATs.
Figure~\ref{fig:orbtyp_sh_rp} shows the median value of $\chi_s$ in 15
bins in $\rperi$ for orbits in the three simulations. The top two
panels show that when simulations are controlled (adiabatic) the
baryons make box and chaotic orbits with smaller pericenter radii less
elongated but at larger pericenter radii these orbits remain elongated
(LATs always tend to be elongated but as shown by
\citetalias{valluri_etal_10}, these too are more elongated when
baryons are absent).  The lower two panels show the elongations of
orbits in g15784. In this simulation all types of orbits are
essentially ``round'' over the entire range in $\rperi$.  Although
this confirms the findings of \citetalias{valluri_etal_10} that orbits
of all three families can become ``round'' to support the shape of the
potential, it is also clear that overall the shapes of all the orbits
are similar and close to axisymmetric at all radii.

\begin{figure*}[t]
\centering \includegraphics[trim=0.pt 0.pt 0.pt 0pt,
  width=0.34\textwidth]{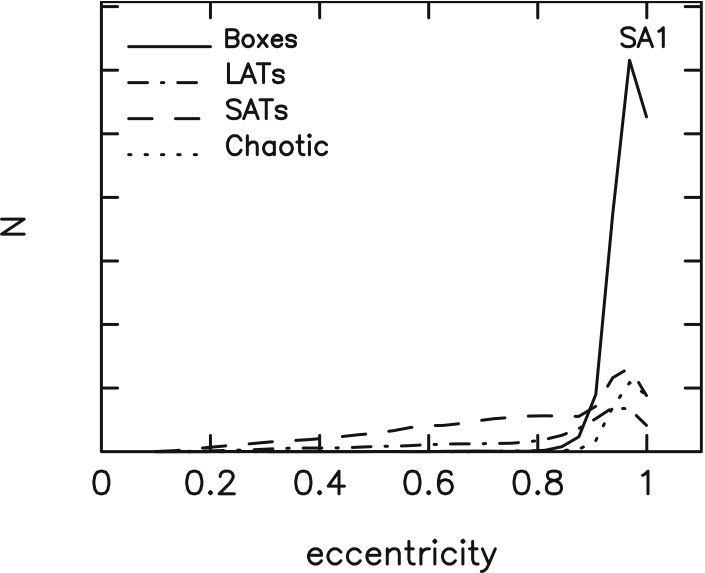}
\centering \includegraphics[trim=0.pt 0.pt 0.pt 0.pt,
  width=0.34\textwidth]{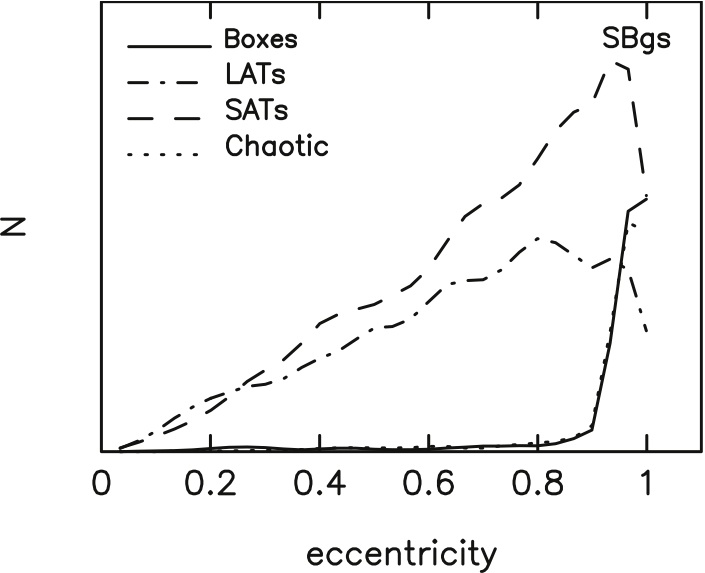}
\centering \includegraphics[trim=0.pt 0.pt 0.pt 0pt,  width=0.34\textwidth]{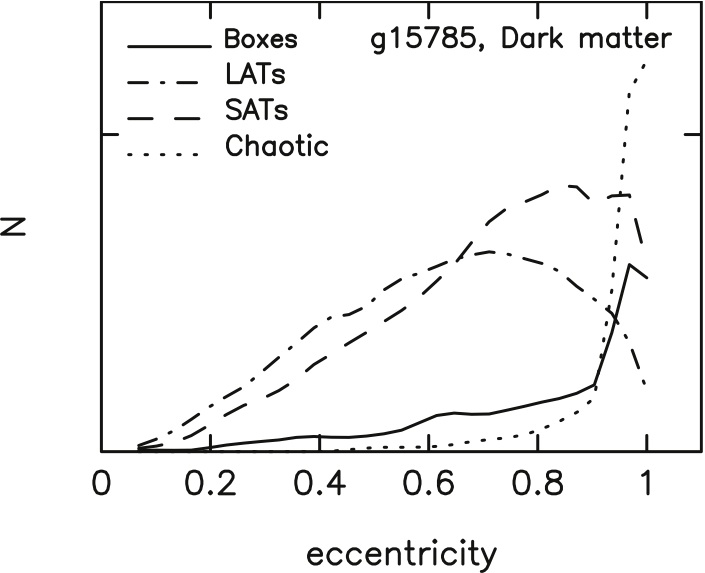}
% Chaotic limit -0.5 , all DM
\centering \includegraphics[trim=0.pt 0.pt 0.pt 0pt, width=0.34\textwidth]{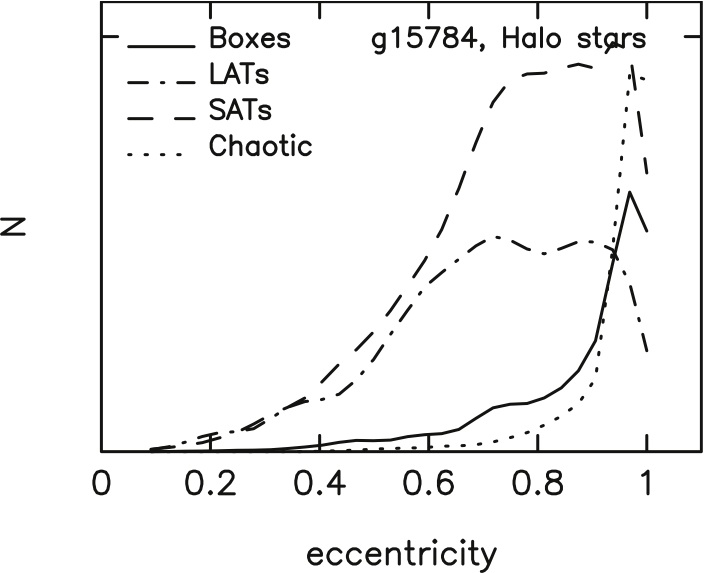}
% Chaotic limit -0.5 , halo stars
%boxes, xtube, ztub, chaotic
%        1063        2616        4333        1128
%  0.11630197      0.28621444      0.47407001      0.12341357   
\caption{Similar to Fig~\ref{fig:orbtyp_peri_r50}: Kernel density distribution of orbit types as a function of their eccentricity.}
\label{fig:orbtyp_ecc}
\end{figure*}

Another commonly used metric for characterizing individual orbits is
the eccentricity parameter $$ e = {\frac{\rapo -
    \rperi}{\rapo+\rperi}}.$$ Strictly speaking, orbital eccentricity
$e$ is defined for a two dimensional orbit in terms of its apocenter
radius $\rapo$ and pericenter radius $\rperi$.  It is important to
emphasize that eccentricity does not measure the degree of
axisymmetry; e.g. a perfectly planar orbit in a spherical potential is
always an axisymmetric rosette, but can have an arbitrary eccentricity
that depends on its angular momentum. At a given energy, orbits with
very high angular momentum have eccentricity approaches zero while
orbits with very low angular momentum have eccentricity  approaches unity. Figure~\ref{fig:orbtyp_ecc} shows the kernel density distributions
of orbital eccentricity for particles in the three simulations. We see
that all four orbit families have larger fractions of orbits with high
eccentricity than with low eccentricity.  Even SATs (dashed curves)
and LATs (dot-dashed curves) are very eccentric implying that they
have fairly low angular momentum.  The absence of low eccentricity box
and chaotic orbits is not unexpected - these orbits are highly radial
(formally having zero time averaged angular momentum) and have $\rperi$ approaches
zero.  The average orbital eccentricities of halo stars in
each orbit family are $\epsilon=0.88$ (boxes), $\epsilon= 0.95$
(chaotic), $\epsilon= 0.76$ (SATs) and $\epsilon=0.72$ (LATs). For
dark matter particles the average orbital eccentricities of each orbit
family are $\epsilon=0.82$ (boxes), $\epsilon= 0.95$ (chaotic),
$\epsilon= 0.71$ (SATs) and $\epsilon=0.63$ (LATs).  These average
orbital eccentricities for the various orbit families are somewhat
larger than the average orbital eccentricities ($\epsilon=0.6$) of
dark matter subhalos and particles in dark-matter-only simulations
\citep{vandenbosch_lewis_lake_stadel_99}.  It is particularly
interesting to reflect on the fact that for g15784 halo stars, even
those on SATs, are more likely to have high eccentricity than low
eccentricity. If a significant fraction of the stellar halo was formed
{\it in situ} in an early disk that was then heated to form the halo,
one would expect this population to have a higher angular momentum and
lower eccentricity than an accreted population. The fact that the
majority of the stellar halo orbits in this global sample have a
relatively high eccentricity reflects the fact that they were
primarily accreted.

The main conclusions from
Figures~\ref{fig:orbtyp_peri_r50} --
\ref{fig:orbtyp_ecc} are:
\begin{itemize}
\item The types of orbits that characterize a halo (either in a
  controlled simulation or in a cosmological simulation) reflect both
  the merger/formation history and the effect of baryons.
\item All four types of orbits (boxes, SATs, LATs and chaotic) can be
  present in significant proportions even when the shape of the
  potential is nearly oblate.
\item All four types of orbits are able to become less elongated (i.e.
  more axisymmetric) to adapt to the nearly oblate shape
\item All four types of orbits have a high fraction of eccentric orbits  reflecting the fact that the halos are products of mergers or radial infall.  
\end{itemize}

\subsection{Phase space distributions: Halo stars versus dark matter}
\label{sec:phasespace}

\begin{figure*}
\includegraphics[trim=0.pt 0.pt 0.pt 0pt,width=0.35\textwidth]{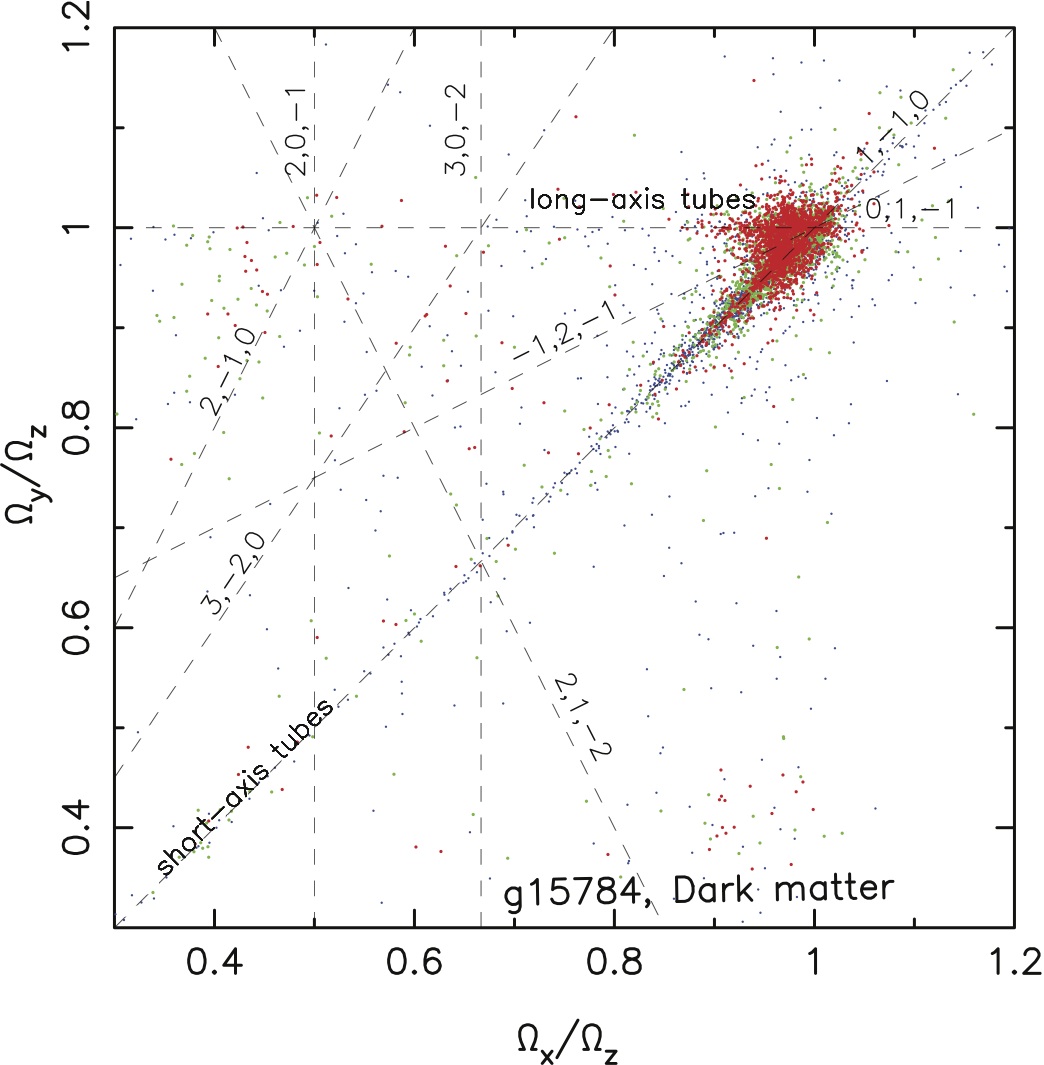}
\centering \includegraphics[trim=0.pt 0.pt 0.pt 0.pt,width=0.34\textwidth]{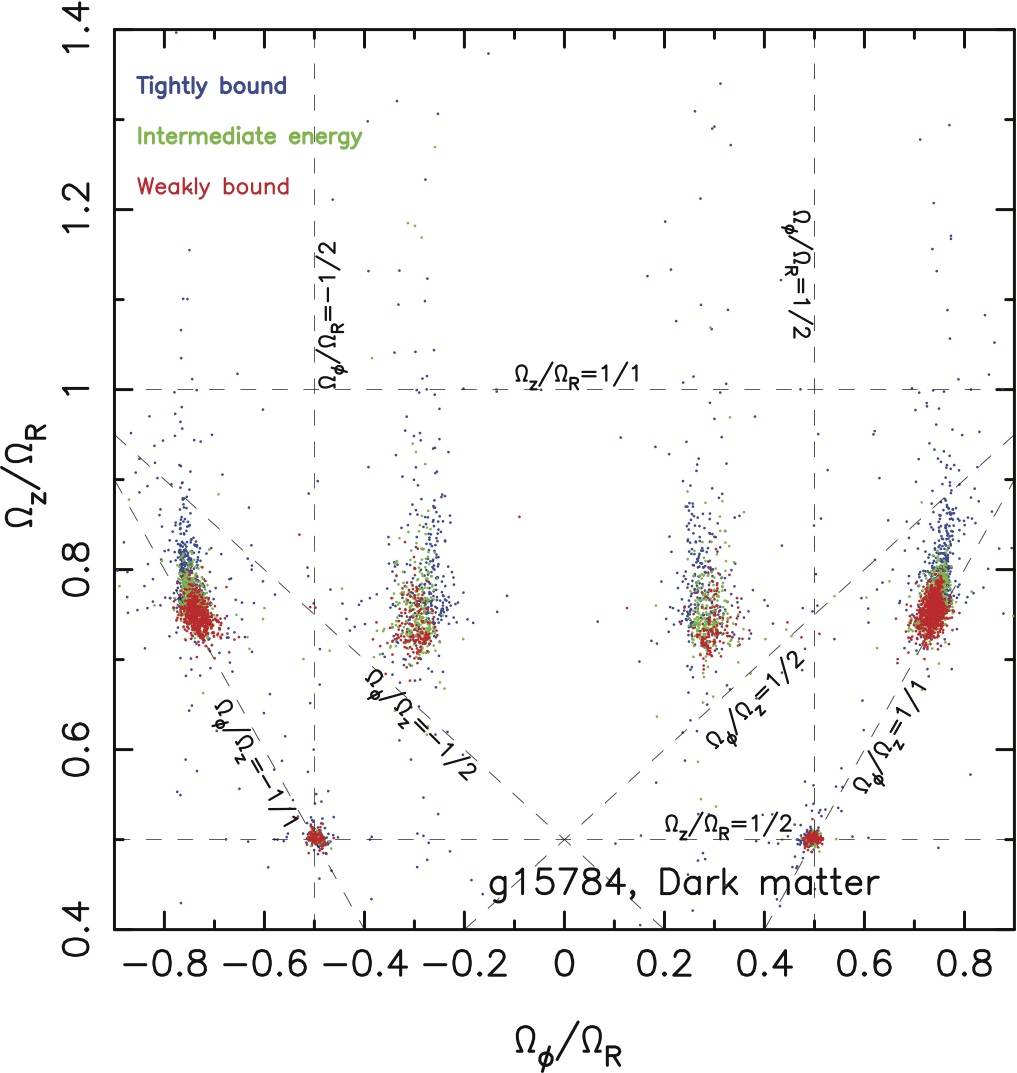}
\includegraphics[trim=0.pt 0.pt 0.pt 0pt,width=0.35\textwidth]{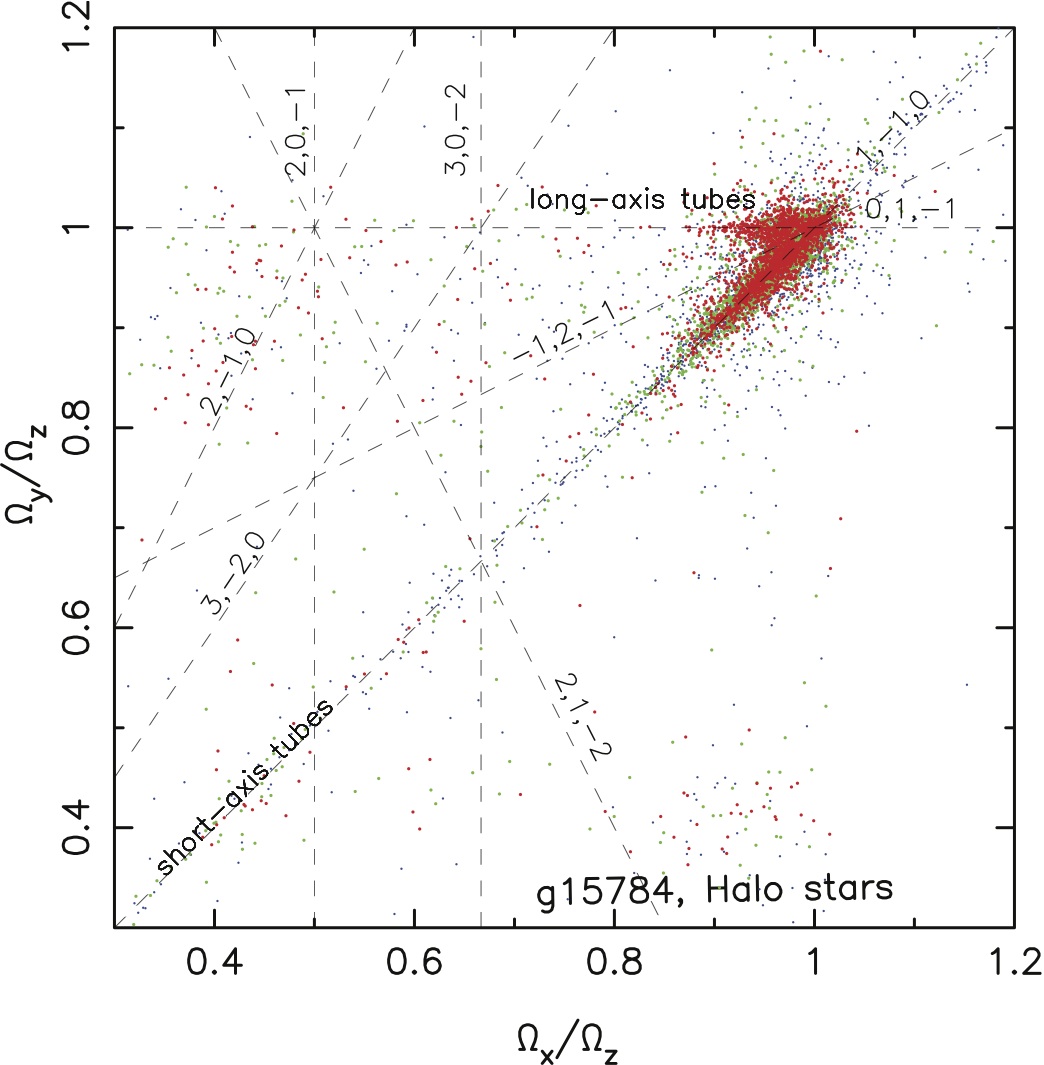}
\centering \includegraphics[trim=0.pt 0.pt 0.pt 0.pt,width=0.34\textwidth]{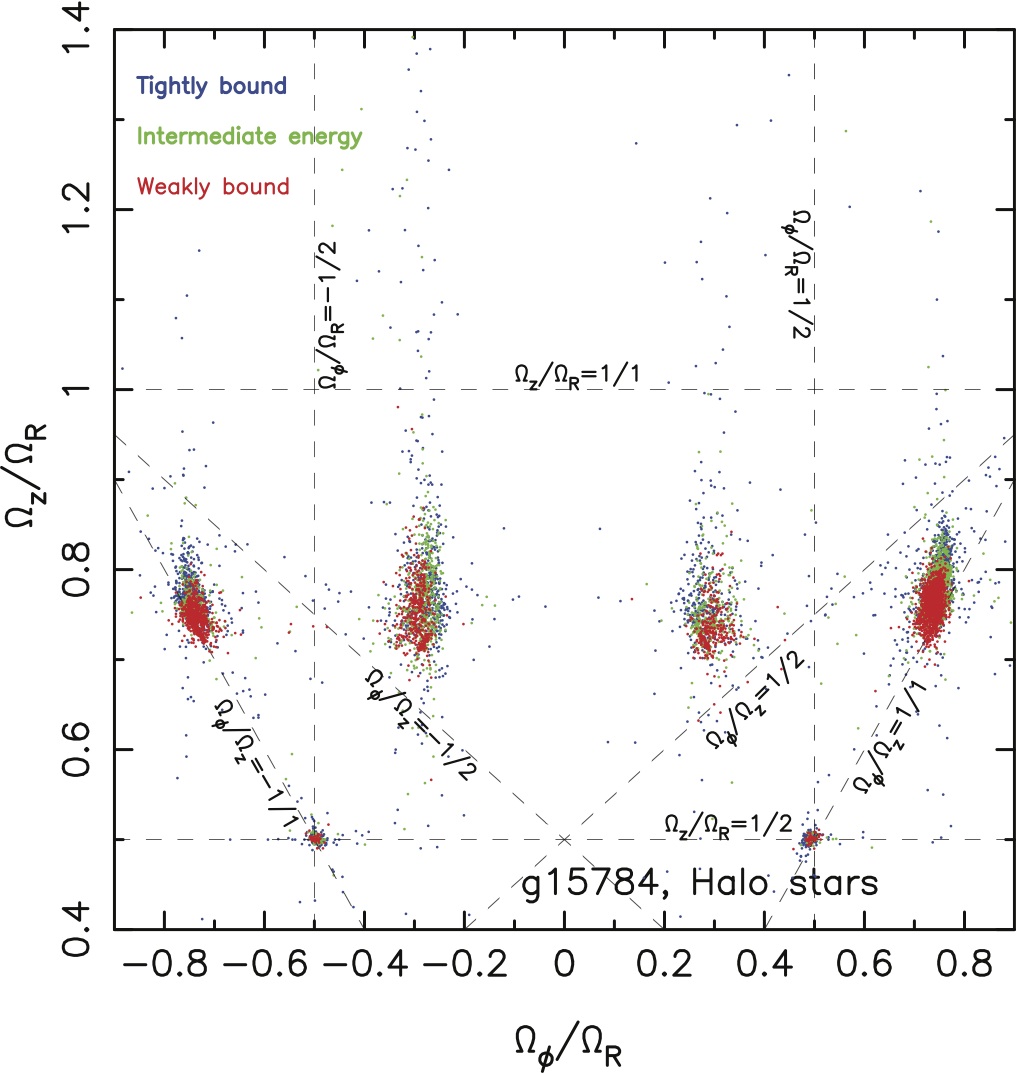}
\caption{Frequency maps for dark matter particles (top row), halo
  stars (bottom row) selected within 50~kpc of the galactic center. Left:
  Frequencies computed in Cartesian coordinates, Right: frequencies
  computed in cylindrical coordinates.Colors represent binding energies of particles and dashed lines mark major resonances (see text).}
\label{fig:cosmo}
\end{figure*}

Since the mass of the stellar halo is only a tiny fraction of the mass
of the dark matter halo, the dynamics of halo stars are determined by
the shape of the dark matter halo, the orbital initial conditions at
the time of accretion, and the evolution of the halo potential since
the epoch of accretion \citep{knebe_etal_05}. All of these factors
also influence the orbital types, orbital shapes and eccentricity
distributions of orbits. The similarities in the distributions of
orbits of dark matter and halo stars with $\rperi$, $\chi_s$ and
$eccentricity$ (Figs. ~\ref{fig:orbtyp_peri_r50} --
\ref{fig:orbtyp_ecc}) show qualitative similarities that are quite
significant, especially in comparison with the orbit populations of
the controlled simulation SA1, which formed from multiple major
mergers. \citet{bryan_etal_12} also found in their analysis of orbits
from the OWLS cosmological simulations that star particles and dark
matter particles have similar orbital distributions as a function of
radius.

\citetalias{valluri_etal_12} showed that a useful way to map the
entire distribution function of a self-consistent system was via
frequency maps.  A frequency map is obtained by plotting the ratios of
fundamental frequencies: in Cartesian coordinates $\Omega_x/\Omega_z$
vs. $\Omega_y/\Omega_z$, or in cylindrical coordinates,
$\Omega_\phi/\Omega_R$ vs.  $\Omega_\phi/\Omega_z$. They showed that
frequency maps give a pictorial representation of the different orbit
families, their relative importance and how they are distributed with
energy.  In the frequency maps in Figure~\ref{fig:cosmo}, each point
represents a single orbit with color representing the orbital binding
energies in 3 bins (blue: the 1/3rd most tightly bound; red: 1/3rd
least tightly bound; green: intermediate energy).

Figure~\ref{fig:cosmo} shows the frequency maps for orbits in galaxy
g15784. The top left panel shows a map (in Cartesian coordinates) for
dark matter particles, while the bottom left shows the same for halo
stars. Short-axis tubes lie along the diagonal $\Omega_x/\Omega_z \sim
\Omega_y/\Omega_z$. The long-axis tube family is clustered along the
horizontal line $\Omega_y/\Omega_z \sim 1$, box and chaotic orbits are
mostly the blue points scattered around the map. The dashed lines mark the major resonances (satisfying conditions like $l\Omega_x+m\Omega_y+n\Omega_z=0$), with
numbers  representing the integers $l,m,n$. In this Cartesian
representation no orbits are seen to be associated with box-orbit
resonances (diagonal lines on the left-side of the graph). The right hand panels of Figure~\ref{fig:cosmo} shows the
frequency maps in cylindrical coordinates for the same sets of orbits.
A prominent resonance is seen at $\Omega_z/\Omega_R =
\Omega_\phi/\Omega_R \sim 0.5$ in orbits of both the halo stars and
dark matter particles. \citetalias{valluri_etal_12} showed that
numerous resonances characterize the phase space of a self-consistent
distribution function evolved adiabatically in controlled
simulations. This resonance is associated with a thin shell resonance
of the SAT family and it is impressive that such resonant orbits
persist despite the hierarchical accretion history of this system. The
frequency map representations of the halo stars and dark matter
particles are very similar to each other both in Cartesian maps (left)
and in the cylindrical maps (right) providing additional evidence that
their phase space distribution functions are similar. This implies
that the phase space distribution function of this sample of stellar
halo orbits probably originated in a manner similar to that of the
dark matter orbits.

\subsection{Galactic archeology: age-metallicity-orbital correlations}
\label{sec:metals}

Since the prescient work of \citet{ELS}, it has been recognized that
correlations between orbital properties of halo stars and their
metallcities and ages can be used to infer the formation history of
the Galaxy. If the orbital integrals of motion of halo stars (e.g.
total energy $E$, the total angular momentum $L$, and angular momentum
about the $z$-axis $L_z$) change only slowly as the potential evolves
and the debris disperses, correlations in integral-of-motion space can
lead to the recovery of the progenitor satellite and also enable
recovery of the underlying potential from tidal debris even in
time-dependent potentials \citep{johnston_etal_96, helmi_dezeeuw_00,
  bullock_johnston_05}.  Searching for correlations in angle-action or
frequency space may improve the accuracy of such recovery
\citep{mcmillan_binney_08, gomez_helmi_10}.  Using semi-analytic
prescriptions to construct stellar populations for accreted halos from
cosmologically motivated pure $N$-body simulations
\citet{font_etal_06b}  showed that combining phase space
information with chemical abundances and stellar ages can aid the
recovery of individual satellites, even when the stars in a satellite
have a small spread in ages and chemical characteristics.  To date
there have been no studies which have tried to identify halo stars
associated with individual satellite progenitors in a {\em
  cosmological hydrodynamical simulation}.

\begin{figure}
\centering 
\includegraphics[trim=0.pt 0.pt 0.pt 0.pt,width=0.45\textwidth]{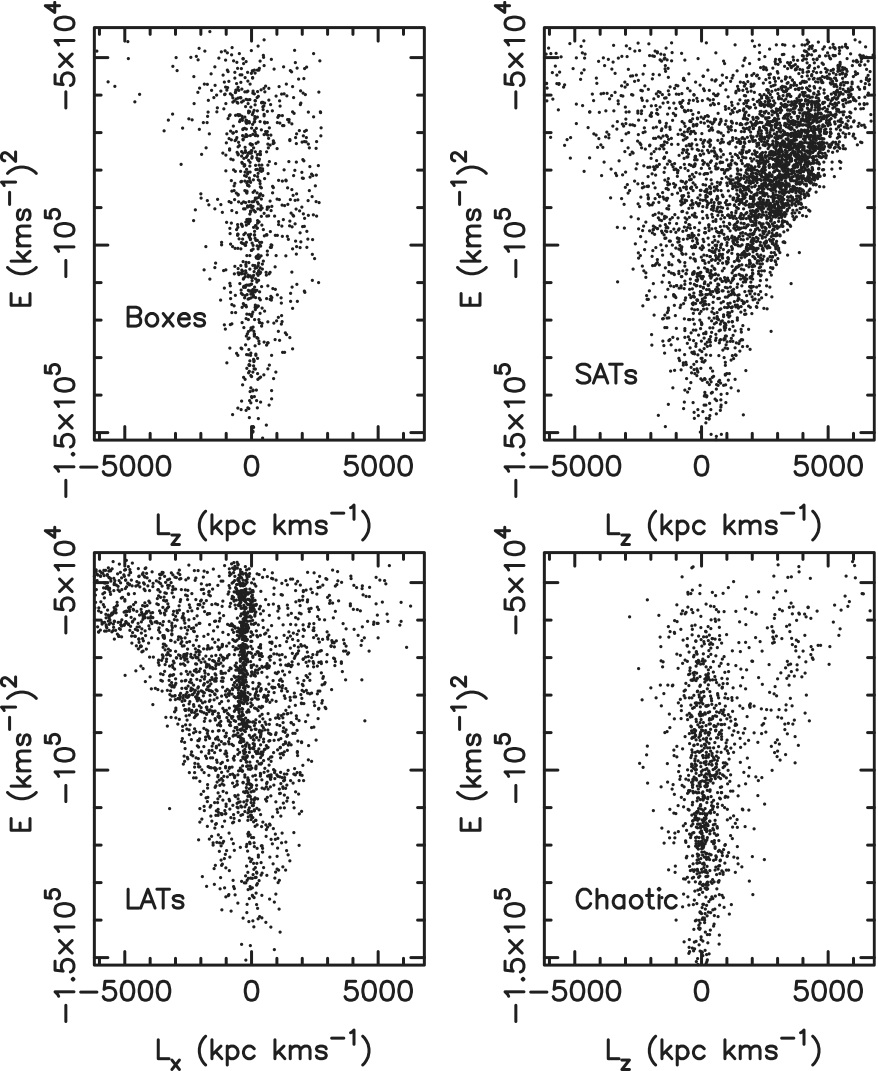}
\caption{Total energy versus one component of angular momentum for the
  global sample of halo stars in g15784. Panels show stars on
  different types of orbits as indicated by the labels. For Box, SAT
  and chaotic orbits ordinate is $L_z$, for LATs (bottom
  left) the ordinate is $L_x$. }
\label{fig:orbtyp_E_J}
\end{figure}

Figure~\ref{fig:orbtyp_E_J} shows orbital energy versus one component
of angular momentum for halo stars from the global sample in each of
the four orbit families studied in Section~\ref{sec:orbtyp}. For box orbits, SATs and chaotic orbits the
abscissa is the angular momentum about the $z$ axis ($L_z$), while for
LATs the abscissa is $L_x$ since $x$ is the axis about which this tube
family rotates.  Points corresponding to orbits in each family are
fairly smoothly distributed and almost no substructure typical of
tidal debris associated with distinct individual satellites
\citep[e.g.][]{helmi_dezeeuw_00} is seen. The $\sim 10^4$ particles
plotted in Figure~\ref{fig:orbtyp_E_J} were selected at random within
a spherical volume of $r_g=50$~kpc. However, even a local sample
selected within a small volume around the location of a fiducial
``sun'' showed no substructure in $E, L, L_z, L_x$ space.  It is
important to emphasize that we only plot $E$ versus $L$ for comparison
with previous work but that in fact {\it none} of the components of
angular momentum are expected to be integrals of motion for {\it any}
of the orbit families in this triaxial
potential. Figure~\ref{fig:orbtyp_E_J} clearly shows that with in this
simulation, the hierarchical growth of the halo, accompanied by
dissipative evolution, have erased all correlations between these
pseudo integrals of motion. It should be noted that our choice of 
orbital initial conditions (randomly chosen within 50~kpc of the galaxy's 
center) is not ideal for searching for coherent tidal debris. Typically for 
such studies a local volume representing the solar neighborhood is 
examined. Our examination of such a volume (i.e. $R_s=4$~kpc yielded too small 
fraction of bona fide halo stars (because the disk in this simulation is quite thick) to perform this exercise. We also searched for substructure in plots of energy vs. metallicity, energy vs. stellar age and angular momentum vs. metallicity and stellar age. None of these plots revealed any detectable substructure that would enable us to identify satellite progenitors.

\begin{figure*}
%omega_3_cyl_ker.f
\centering 
\includegraphics[trim=0.pt 0.pt 0.pt 0.pt,width=0.75\textwidth]{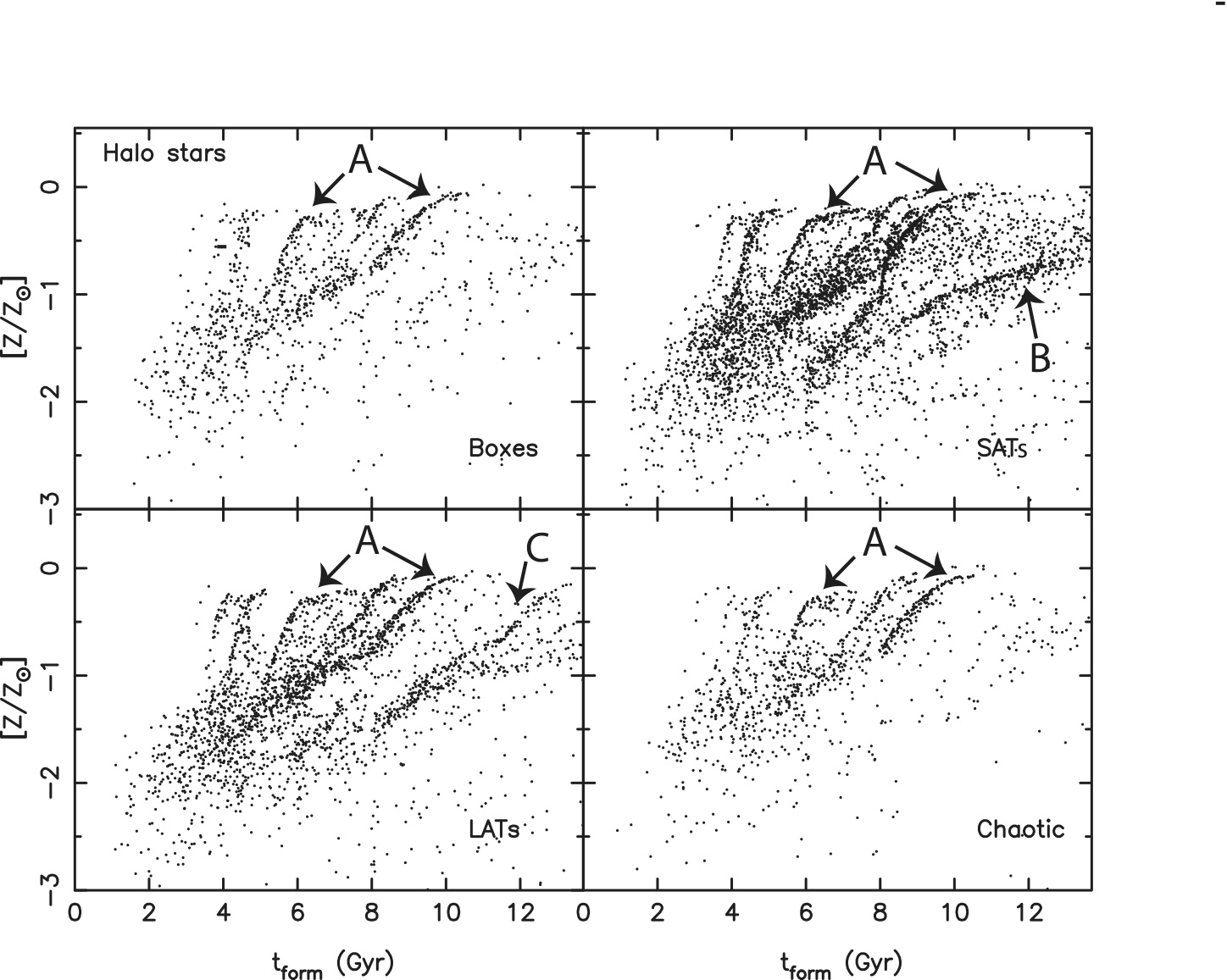}
\caption{Metallicity (in solar units) versus formation time ($\tform$) for halo stars in the global sample. Panels show stars on different types of orbits as indicated by labels.  See text for discussion of labels ``A'', ``B'' and ``C''.}
\label{fig:orbtyp_Met_tform}
\end{figure*}

Bailin et al. (2013, in preparation) analyze the age-metallicity
relation for all stars (halo and disk) in the MUGS cosmological disk
galaxy simulation g15784. They find that while {\it the most}
metal-poor stars in the galaxy were formed at the {\it earliest}
times, the ISM is rapidly enriched and hence the formation of highly
metal enriched stars follows very quickly thereafter.  Bailin et
al. show that when $[Fe/H]$ is plotted versus formation time, numerous tight ``stream-like'' features which correspond to continuous star formation within individual
galactic subunits, are seen.  A study of the relative spatial,
metallicity and age distributions of stars that were accreted onto g15784 and
stars that formed in the disk and were subsequently kicked into the halo will be presented in Bailin et al. Here we focus only on $10^4$ halo stars whose orbits were analyzed in
previous sections, and examine how their orbital characteristics
depend on their metallicities and ages.

In Figure~\ref{fig:orbtyp_Met_tform} we plot total metallicity (in
solar units) for stars in the halo orbit sample studied in previous
sections, versus their formation time in the simulation
($\tform$). The formation time is scaled such that stars that form at
the present time ($z=0$) have $\tform= 13.7$~Gyr. The panels shows
stars on different types of orbits: boxes, SATs, LATs and chaotic
orbits as indicated by the labels. Each point represents a star
particle from the simulation.

Most striking are the thin stream-like features (identified by Bailin
et al.) which arise from progressive enrichment and star-formation
inside individual sub-galactic units.  Note that even though we have
{\em randomly} selected $\sim 10^4$ halo stars within 50~kpc, with no
other constraints other that they have $\rapo> 5$~kpc and $\zmax
>3$~kpc the thin streams associated with individual progenitors are
clearly visible. Star formation appears to stop at some point in all
the streams.  While more detailed analysis with all the MUGS snapshots
are needed to determine the precise reason for this, a plausible
interpretation is that star formation in a subhalo can continue for some time after it is accreted by the galaxy but stops once it is tidally disrupted.  Consequently the end of the ``stream'' serves as a proxy for when it is significantly disrupted by tidal forces from the main galaxy.

The thin stream-like features which were disrupted early ($\tform <
8-10$~Gyr) appear in all four panels indicating that at $z=0$, stars
associated with a single progenitor satellite traverse the halo on all
four types of orbits. For example the features labeled ``A'' consists
of stars with a wide range of metallicity and $\tform$\, and are
clearly seen in four panels. This implies that these two galactic
sub-units were completely disrupted by tidal forces and then scattered
onto different types of orbits by  mixing processes. 

We believe this is probably ``chaotic mixing'' rather than ``phase mixing'' since the
latter conserves orbital integrals of motion in a static potential and
changes them adiabatically in a slowly varying one.  Phase mixing is
not capable of changing the orbital type, but chaotic mixing is
\citep{merritt_valluri_96}. Therefore it is unlikely that the phase
space coordinates of stars (at $z=0$) in the two units labeled ``A''
can be used to fully reconstruct the progenitor or its orbital
properties at the time it was accreted. For dynamically older streams
like ``A'' the integrals of motion, especially energy, are expected to
change significantly due to the change in the halo potential due to
mass accretion, both dark \citep{knebe_etal_05} and baryonic
(\citetalias{valluri_etal_10}).  Additionally chaotic mixing in the
triaxial halo causes tidal streams to disperse more rapidly
\citep{vogelsberger_etal_08}.

In contrast the $t_{form}$-metallicity feature labeled ``B'' is most clearly seen in the top right panel (SATs). Its stars also span a range of metallicities
($-1.5< [Z/Z_\odot]<-0.5$) and formation epochs (6-13.5~Gyr) however
most of the stars of this satellite are on SATs (although a faint hint
of it is seen in the LATs panel) while there is no distinct counterpart in the box or chaotic orbit panels\footnote{although admittedly this could be in part due to inadequate sampling.}. This indicates that the orbits of this satellite
are {\it not mixed}, and hence the progenitor's phase space
coordinates and the halo potential can probably be extracted by
backward evolution of stellar orbits. The stream labeled ``C'' in the
bottom left panel consists of stars on LATs and is also only seen in
this panel.  The near-exclusivity in orbit-type for streams ``B'' and
``C'' implies that the orbits of these stars have not experienced much
mixing (probably because they were accreted/tidally disrupted more recently).
Figure~\ref{fig:orbtyp_Met_tform} shows that the thin steam-like
features have stars with a wide range of metallicities and star formation occurs over an extended period of time.  We see that stars with a
range of metallicities and ages that were once associated with the
same progenitor are easy to identify because they form thin features
in metallicity-formation time plots despite sometimes having very
different orbital characteristics.

Our preliminary analysis of the age-metallicity signatures of our selected halo stars in Figure~\ref{fig:orbtyp_Met_tform}  shows that satellites leave the stars in long and short-axis tube orbits as they accrete.  Satellites that accreted >5 Gyr ago have chaotically mixed these stars into chaotic and box orbits, but more recently accreted satellites have not yet become well mixed.  Such chaotic mixing of early accreted satellites is consistent with previous studies of satellite accretion \citep{bullock_johnston_05,font_etal_06b,gomez_helmi_10}.

  %However significantly more information about the accretion and star formation history of the Milky Way halo will be available from a combination of stellar ages, metallicities and phase space coordinates.
It is important to note that star formation appears to continue in this halo population till quite late times --- well beyond what is expected based on observations of the stellar halos of the Milky Way and M31. This is in part due an artifact of the detailed sub-grid physics in the simulations: the cooling of gas in to dark matter subhalos results in their becoming more massive than in the real Universe. This also allows subhalos in these simulations to continue to form stars after they are accreted by the main galaxy. This (presumed) artifact is not unique to our simulations and has been previously observed \citep{tissera_etal_12,tissera_etal_13}. \citet{tissera_etal_13} refer to this component of the stellar halo as ``endo-debris'', and distinguish it from  ``debris'' stars which were formed in a satellite before it was accreted. There is tentative evidence that such a population has also been discovered in the Milky Way halo by \citet{sheffield_etal_12} who have identified a population of halo stars with low [Fe/H] but high [$\alpha$/Fe], which they argue could have been formed {\it in situ} in the halo (i.e. inside satellites after they were accreted). \citet{tissera_etal_13} also suggest that one  population of Carbon Enhanced Metal Poor stars found by \citet{carollo_etal_12} could be such ``endo-debris'' stars. If star formation in satellites terminates earlier than in these simulations, narrow features would not extend to late times, but would still remain detectable at early times as in Figure~\ref{fig:orbtyp_Met_tform}.  

Together Figures~\ref{fig:orbtyp_E_J} and \ref{fig:orbtyp_Met_tform} imply that the tidal debris in cosmological hydrodynamical simulations experiences significantly more chaotic
evolution than in collisionless simulations, making it much harder to identify individual progenitors using phase space coordinates alone. But the identification of distinct progenitors is quite easy in age-metallicity plots, and such plots when combined with orbital information can be used to probe the accretion history of the halo. Current cosmological hydrodynamic simulations, including the one studied here, do not have adequate resolution to identify the types of stellar streams observed in the Milky Way and M31. The study of  the correlations between kinematics, metallicities, and spatial distributions of halo stars and their formation history will become increasingly important in the future, as the resolution of simulations and the baryonic physics prescriptions improve.

\section{Summary and Conclusions }
\label{sec:discuss}

We have analyzed the orbital properties of halo stars and dark matter
particles from a disk galaxy forming in a cosmological hydrodynamical
simulation from the MUGS project. The simulated disk galaxy we have
studied is a reasonably good Milky Way analog, although it has a
central bulge and satellite galaxies that are too massive.  In this
galaxy the shape of the dark matter and stellar halos vary with radius
due to the central condensation of baryons: oblate within about
20~kpc, and mildly triaxial at intermediate and large radii,
consistent with previous work \citep{dubinski_carlberg_91,
  kazantzidis_etal_04shapes, deb_etal_08,tissera_etal_10,
  kazantzidis_etal_10, zemp_etal_12}. The stellar halo is slightly
more oblate than the dark matter halo at all radii. We find that
although the inner halo region -- where the disk dominates -- is
oblate, it is dominated by box orbits and chaotic orbits and less than
30\% of orbits in the inner oblate region are SATs.  This is contrary
to the general expectation that as an ellipsoidal figure becomes more
oblate, its distribution function is increasingly dominated by SATs
\citep{gerhard_binney_85,merritt_valluri_96,BT}. In the outer regions,
where the potential is more elongated, SATs and LATs appear in roughly
equal proportions and box orbits and chaotic orbits are insignificant.
{\em Thus in no case do we find evidence that oblate regions of the
  potentials are supported mainly by SATs as is generally assumed.}
Our findings confirm our previous work with controlled simulations
(D08, \citetalias{valluri_etal_10}).

The phase space distributions of halo stars and dark matter particles
(as characterized by orbit populations as a function of radius and
eccentricity) in the cosmological hydrodynamical simulations are very
similar to each other.  Although orbits of both types have a wide range of orbital eccentricities, the majority (even of those on SATs)  have high eccentricity,
indicative of their largely accreted origin; the distributions of orbital types with pericenter radius, orbital elongation parameter and orbital eccentricity differ only slightly.
Orbits of all four families have three dimensional shapes (as measured by the elongation
parameter $\chi_s$) that are close to axisymmetric, enabling them to
self-consistently generate the nearly oblate shape of the inner halo.   Where there are differences between the orbits of star and dark matter particles, they
are clearly attributable to the differences in their spatial
distributions.  For instance, since the dark matter distribution is
more extended and more triaxial than the stellar distribution, it is
dominated by LATs in the outer regions.  The less extended stellar
halo is also more oblate and hence SATs dominate even beyond 20~kpc.
Similarities in the orbital phase space properties (and distribution
with energy) are also seen in the frequency maps.  

The similarities between the orbital structures of the stellar halo and dark matter
halo suggest that {\em both components share a common dynamical origin,
probably accretion}, although it is possible that some of the
similarities arise because chaotic scattering of orbits of both types
of particles, in the fully cosmological simulation. The relatively
large fraction of chaotic orbits may not be representative of real
galaxies and is most likely a consequence of  the
presence of the massive central bulge and subhalos which produces some
scattering (D08), and the ``freezing'' of irregularities like small
subhalos and tidal features that accentuate the scattering of resonant
orbits.

Several previous studies have focused on uncovering tidal streams from
the orbital phase space coordinates of halo stars in dissipationless
$N$-body simulations. In this fully cosmological hydrodynamical
simulation we find that energy and angular momentum (or other quantities resembling integrals of motion such as orbital actions) --- the variables
traditionally used to identify tidal debris in idealized collisionless
simulations of tidal disruption \citep{helmi_dezeeuw_00} --- evolve so significantly that they can no longer be used (exclusively) for identifying substructure in phase space. The importance of chaotic mixing in such simulations is evidently larger than in the controlled N-body simulations studied previously.    

We  show that in plots of metallicity versus formation time (or
stellar age), thin stream like features can be associated with
individual galactic satellites in which star formation occurs
continuously over a long period of time before it is tidally disrupted in the main galaxy's halo (Fig.~\ref{fig:orbtyp_Met_tform}). Consequently stars associated with a single disrupted satellite are neither of single metallicity nor do they have a narrow range of stellar ages \citep[as also found by][]{font_etal_06b}.  For older disruption events matched stream-like features are found on all four families of orbits implying that they have experienced chaotic mixing. However, younger features are typically associated with a single orbit family. 

The results of this paper demonstrate the value of orbital analysis for probing dynamical structure and formation history. Similar studies of simulated triaxial
galaxies and dark matter halos show that systems that form via dissipationless ``dry
mergers'' contain primarily box orbits and long-axis tubes \citep{barnes_96,valluri_etal_10} while tube orbits tend to be
produced by dissipation in ``wet mergers'' \citep{hoffman_etal_10}. Conversely if a system is dominated by box orbits or chaotic orbits one can infer that dissipationless
formation such as a multi-stage merger or that scattering by a dense
center has been important in the evolution of the system.  

It is important to point out that the MUGS simulations used in this
paper have recently been superseded by several simulations in which
new feedback prescriptions are able to produce either bulgeless disks
or disks with significantly larger disk/total ratios
\citep{governato_etal_10, guedes_etal_11, zemp_etal_12,
  stinson_etal_12}. Future generations of simulations will continue to
improve the match between observations and simulations.  It is
important to carry out similar analyses in these improved simulations
to assess how specific implementations of baryonic physics alter the
star formation rates and enrichment of subhalos that make up the
stellar halo.
 
In coming years phase space coordinates, elemental abundances and accurate isochronal ages for many halo stars will be obtained with Gaia and future  wide-field, high-precision synoptic surveys, such as LSST. Surveys such as Gaia have been designed to yield data with observational errors of 10\% or less for stars within 10~kpc of the sun. Several ground based surveys will go even deeper.  In a future paper we will
consider the effects of selection bias and observational errors on the
phase space coordinates, stellar ages, and elemental abundances on the
recovery of the orbital structure of the stellar halo.

%The orbital properties of cosmological and controlled halo simulations studied in the paper show that different formation histories are imprinted in the distributions of orbital types and can be used to uncover clues to this evolutionary history. This idea is not new: it dates back to \citep{ELS} and is a major motivation for the large number of upcoming surveys designed to obtain 6D phase space coordinates, ages and elemental abundances for billions of stars.

\acknowledgements MV is supported by NSF grant AST-0908346 and
University of Michigan's Elizabeth Crosby grant. She would like to acknowledge the contribution of David Thompson to the analysis of halo shapes. MV \& JB thank
Eric Bell,  Facundo Gomez,  Antonela Monachesi  and Colin Slater, for their energetic discussions of stellar halo issues.

\bibliography{MUGS}

\end{document}